\documentclass[aps,superscriptaddress,nofootinbib]{revtex4}
\pdfoutput=1

\usepackage{graphicx}
\usepackage{slashed}
\usepackage{amssymb,amsmath,amsbsy}

\usepackage{psfrag}

\usepackage{epsfig}

\def\be{\begin{equation}}
\def\ee{\end{equation}}

\newcommand{\bea}{\begin{eqnarray}}
\newcommand{\eea}{\end{eqnarray}}
\newcommand{\no}{\nonumber\\}

\begin{document}

\title{Are There Hidden Scalars in LHC Higgs Results? }
\author{A.~Arhrib}
\email{aarhrib@ictp.it}
\affiliation{D\'epartement de Math\'ematiques, Facult\'e des Sciences et Techniques, Tanger, Morocco}
\author{P.M.~Ferreira}
    \email[E-mail: ]{ferreira@cii.fc.ul.pt}
\affiliation{Instituto Superior de Engenharia de Lisboa - ISEL,
	1959-007 Lisboa, Portugal}
\affiliation{Centro de F\'{\i}sica Te\'{o}rica e Computacional,
    Faculdade de Ci\^{e}ncias,
    Universidade de Lisboa,
    Av.\ Prof.\ Gama Pinto 2,
    1649-003 Lisboa, Portugal}
\author{Rui Santos}
    \email[E-mail: ]{rsantos@cii.fc.ul.pt}
\affiliation{Instituto Superior de Engenharia de Lisboa - ISEL,
	1959-007 Lisboa, Portugal}
\affiliation{Centro de F\'{\i}sica Te\'{o}rica e Computacional,
    Faculdade de Ci\^{e}ncias,
    Universidade de Lisboa,
    Av.\ Prof.\ Gama Pinto 2,
    1649-003 Lisboa, Portugal}

\date{\today}

\begin{abstract}
The Higgs boson recently discovered at the Large Hadron Collider has
shown to have couplings to the remaining  particles well within what
is predicted by the Standard Model. The search for other new heavy scalar states
has so far revealed to be fruitless, imposing constraints on
the existence of new scalar particles. However, it is still possible that
any existing heavy scalars would preferentially decay to final states involving
the light Higgs boson thus evading the current LHC bounds on heavy scalar states.
Moreover, decays of the heavy scalars could increase the number of light Higgs bosons
being produced. Since the number of light Higgs bosons decaying to Standard Model
particles is within the predicted range, this could mean that part of the light Higgs bosons
could have their origin in heavy scalar decays. This situation would occur if the light
Higgs couplings to Standard Model particles were reduced
by a concomitant amount. Using a very simple extension of the SM - the two-Higgs double model
- we show that in fact we could already be observing the effect of the heavy scalar states even
if all results related to the Higgs are in excellent agreement with the Standard Model predictions.
\end{abstract}


\maketitle

The recent LHC discovery of the Higgs boson~\cite{:2012gk,:2012gu}
completed the particle spectrum predicted by the Standard Model (SM). We are
finally able to directly probe the SM's scalar sector, and
eventual discrepancies with SM expectations will likely be signs of new physics
or new particles. One of the simplest extensions of the SM is the two-Higgs
doublet model (2HDM), proposed by Lee in 1973~\cite{Lee:1973iz} as a means to
explain matter-antimatter asymmetry. For a recent thorough review of the 2HDM,
see~\cite{Branco:2011iw}.

The model has a vast and interesting phenomenology. It boasts not one but {\em five}
scalar states - two CP-even scalars (the lightest $h$ and the heaviest $H$), a
pseudoscalar ($A$) and a pair of charged scalars ($H^\pm$). The vacuum structure of the model
is much richer than the SM's: for instance, the 2HDM scalar potential can have minima
which spontaneously break CP, thus aiding in explaining the abundance of matter over
antimatter in the universe. But minima which break electric charge conservation are
also possible, as are vacua which preserve electric charge conservation and CP while breaking the
electroweak symmetry but are tree-level metastable~\cite{Ivanov:2006yq,Ivanov:2007de,Barroso:2007rr}.
Analytical conditions to avoid those dangerous vacua have been obtained and are very
simple to implement~\cite{Ferreira:2004yd,Barroso:2005sm,Barroso:2012mj,Barroso:2013awa}.

The Yukawa sector of the 2HDM is also quite interesting: {\em a priori} it exhibits
tree-level FCNC, but these can either be eliminated through a discrete $Z_2$
symmetry~\cite{Glashow:1976nt,Paschos:1976ay} or be made ``naturally small", by means
of an appropriate global symmetry~\cite{Branco:1996bq}.
Also, the judicious application of the $Z_2$ symmetry can yield a version of the
model - the ``inert" 2HDM - which has candidates for dark matter~\cite{inert}.

Experimental confirmation of the 2HDM will obviously necessitate the discovery of its
extra scalars: out of the five states that the model predicts one has just been
discovered, and we take it to be the lightest CP-even scalar. It is unlikely, though
not impossible~\cite{heavyh}, that the LHC-observed scalar is actually the heaviest of
the CP-even states, $H$, and the observation of the Higgs decays into $W^+W^-$ and
$ZZ$ excludes the possibility of it being the pseudoscalar $A$ (though the
possibility that the LHC discovery corresponds to a mixed scalar-pseudoscalar particle
is not yet ruled out~\cite{mix}).

Assuming therefore that the scalar $h$ has been observed, what can we already say, with
current LHC data, about the remaining 2HDM scalars? What does the non-observation of $H$,
$A$ and $H^\pm$ to date tell us? And how much of parameter space of the 2HDM is curtailed
by not having found these other scalars? Similar questions were recently analysed in
ref.~\cite{Chen:2013rba}.

In this paper we will perform a thorough examination of these questions and show that, though
current LHC results already exclude a sizeable portion of 2HDM parameter space, there is still
ample room to allow for the existence of extra scalars with relatively low masses (below 600
GeV). And we will study also an intriguing possibility - that heavier scalars like $H$ may
already be indirectly observed at the LHC, not through their decays into gauge bosons or
fermions, but rather through their decays into $h$. In fact, such decays (for instance,
$H \rightarrow hh$, $A \rightarrow Zh$ or $H^\pm \rightarrow W^\pm h$) can, for ordinary
values of the parameters of the model, be the dominant contribution to the branching ratios
of the heavier scalars. If that were the case then, for instance, $H$ would be invisible in decays
to gauge bosons or fermions, and its presence could well already be detected through its
contribution to $h$ production. The lightest CP-even scalar $h$ could therefore be produced
as a result of a chain of production and decays of $H$, $A$ and/or $H^\pm$, and we call this process
{\em chain Higgs production}.
This seemingly exotic scenario can occur without any fine tuning and, as we
will see, is in agreement with the existing data.

In section~\ref{sec:model} we review basic notation and facts of the 2HDM, describing
the constraints, both theoretical and experimental, that the model is subject to. And we explain
the scan of parameter space we will be performing, trying to cover as vast a range as
possible. In section~\ref{sec:sigma} we detail the calculation of cross section values needed for
this work - not only for $h$ production but also that of $H$, $A$ and $H^\pm$, for a wide
range of values of their masses and other variables. In section~\ref{sec:heav} we introduce
the variables which we compare with LHC data and show how that data already excludes large chunks
of 2HDM parameters due to non-observation of the heavier states. Nonetheless, there is plenty
of room left to accommodate the existence of new scalars. In section~\ref{sec:chain} we treat
the possibility of chain Higgs production through decays of the heavy scalar states. We will
show that there may be a sizeable contribution to the observed Higgs signals at the LHC coming from these
processes, which would mean we already may be, indirectly, observing the remainder of the 2HDM
scalar spectrum. We will briefly discuss what could we do to probe this possibility, and present
our conclusions in section~\ref{sec:conc}.

\section{The Two Higgs Doublet Model}
\label{sec:model}

The 2HDM has the same fermion and gauge particle content of the SM, and an extra
$SU(2)\times U(1)$ doublet $\Phi_2$, with hypercharge $+1$ like the first doublet
$\Phi_1$. The most general 2HDM scalar potential has 11 independent parameters.
In the most used version of the model one imposes a discrete $Z_2$ symmetry, $\Phi_1
\rightarrow \Phi_1$ and $\Phi_2\rightarrow -\Phi_2$, which eliminates four parameters.
We will softly break this $Z_2$ symmetry and thus the scalar potential is written as
\bea
V &=&
m_{11}^2 |\Phi_1|^2
+ m_{22}^2 |\Phi_2|^2
-  m_{12}^2 \left(\Phi_1^\dagger \Phi_2 + h.c. \right)
\no & &
+ \frac{1}{2} \lambda_1 |\Phi_1|^4
+ \frac{1}{2} \lambda_2 |\Phi_2|^4
+ \lambda_3 |\Phi_1|^2 |\Phi_2|^2
+ \lambda_4 |\Phi_1^\dagger\Phi_2|^2
+ \frac{1}{2} \lambda_5 \left[\left( \Phi_1^\dagger\Phi_2 \right)^2
+ h.c. \right],
\label{eq:pot}
\eea
with all parameters real~\footnote{Another possibility would be to take a complex soft
breaking term $m_{12}^2$, yielding the {\em complex 2HDM}. This model has many interesting
features~\cite{chdm, mix} but will not be studied in the current work}. When both doublets
acquire vacuum expectation values (vevs),
$\langle \Phi_1 \rangle = v_1$, $\langle \Phi_2 \rangle = v_2$, electroweak
symmetry breaking occurs. Looking at
eq.~\eqref{eq:pot} we see the potential has 8 independent real parameters. Usually, though, we
take as independent parameters the set of scalar masses $m_h$, $m_H$, $m_A$ and $m_{H^\pm}$, the
sum of the squared vevs, $v^2\,=\,v_1^2 + v_2^2 \,=\, (246$ GeV$)^2$, the angle $\beta$
defined through the vev ratio, $\tan\beta = v_2/v_1$, the mixing angle $\alpha$ of the CP-even
$2\times 2$ mass matrix and the soft breaking parameter $m_{12}^2$. Since the lightest Higgs mass
has been measured to be $\simeq 125$ GeV, the model has therefore 6 unknown parameters.

The coupling of the scalars to the fermions is also affected by the $Z_2$ symmetry. But the extension
of that symmetry to the fermion fields, {\em i.e.} how they transform under $Z_2$, is not uniquely
defined, thus giving rise to different versions of the 2HDM Lagrangian, with different phenomenologies.
Model Type I corresponds to all fermion fields swapping sign under $Z_2$ and thus coupling only to
the doublet $\Phi_2$. In model Type II the fermion transformation laws are such that charged leptons
and down-type quarks couple to $\Phi_1$ and up-type quarks couple to $\Phi_2$~\footnote{There are other
versions of the model (X and Y), differing in the specific couplings to leptons~\cite{othermodels}, but we
will not consider them here.}.

In all that follows, when we speak of parameter scans, we have taken $m_h = 125$ GeV,
$127\leq m_H \leq 900$ GeV, $100\leq m_A \leq 900$ GeV,
$100\leq m_{H^\pm} \leq 900$ GeV~\footnote{For model type-II, due to a strong constraint stemming
from B-physics (to be discussed shortly), we generate only points with $m_{H^\pm} \geq 360$ GeV.},
$1 \leq \tan\beta \leq 30$, $-\pi/2 \leq \alpha\leq \pi/2$ and
$|m_{12}^2| \lesssim 4\times10^5$ GeV$^2$.

\subsection{Constraints on 2HDM parameters}

Even before the Higgs discovery, there was plenty one could say about the allowed values for
the parameters of the model. For starters, they have to be such that electroweak symmetry breaking
occurs. And one must ensure that the potential~\eqref{eq:pot} is bounded from below~\cite{vac1}, which
imposes a set of positivity constraints on the quartic couplings:
\bea
\lambda_1 > 0 & , &  \lambda_2 > 0 \; ,\nonumber \\
\lambda_3 > -\sqrt{\lambda_1 \lambda_2} & , &
\lambda_3 + \lambda_4 - |\lambda_5| > -\sqrt{\lambda_1 \lambda_2} \;,
\label{eq:bfb}
\eea
Also, to make sure that unitarity is preserved and the couplings remain perturbative, there
is a second set of constraints imposed on the quartic couplings~\cite{unitarity}. And if one wishes
to be sure that the vacuum of the potential is the global minimum of the model, one ought
to impose the following condition~\cite{Barroso:2012mj,Barroso:2013awa},
\be
m^2_{12} (\sqrt{\lambda_2} m^2_{11} - \sqrt{\lambda_1} m^2_{22})
(\sqrt{\lambda_2}\tan\beta - \sqrt{\lambda_1})\,>\,0.
\label{eq:disc}
\ee
The precision electroweak constraints
are also taken into account~\cite{Peskin:1991sw, STHiggs}.

Regarding the experimental bounds, the most relevant are the ones imposed
on the charged Higgs mass and on the $\tan \beta$ parameter coming from B-physics,
LEP and from the most recent LHC data. The LEP experiments have set a lower
limit on the mass of the charged Higgs boson of 80 GeV  at 95\% C.L., assuming
$BR(H^+ \to \tau^+ \nu) + BR(H^+ \to c \bar s) +  BR(H^+ \to A W^+) =1$~\cite{LEP2013}
with the process $e^+ e^- \to H^+ H^-$. The bound
is increased to 94 GeV if $BR(H^+ \to \tau^+ \nu)  =1$~\cite{LEP2013}.
Using the constraints~\cite{BB} coming from the measurements of $R_b$, $B_q \bar{B_q}$
mixing and also from $B\to X_s \gamma$~\cite{BB2}, we conclude that values of $\tan \beta$
smaller than $O (1)$ together with  a charged Higgs with a mass below $O (100$ GeV$)$ are disallowed
for all Yukawa-types. Moreover, for the particular cases of models type II and Y, $B\to X_s \gamma$~\cite{BB2}
imposes a lower limit on the charged Higgs
mass of about $360$ GeV.

The LHC, through both the ATLAS~\cite{ATLASICHEP} and the CMS~\cite{CMSICHEP} collaborations,
also places bounds on the $(\tan \beta, m_{H^\pm}) $ for light charged Higgs with
 $pp \to t \bar{t} \to b \bar{b} W^\mp H^\pm (\to \tau \nu)$.
Finally, we have considered the LEP bounds on the neutral Higgs~\cite{mssmhiggs}.

We end up with a note about the anomaly observed in the rates $R (D)$ and $R(D^*)$,
with $R(D) = (\overline{B}\to D \tau^-\overline{\nu}_\tau)/(\overline{B}\to D l^-\overline{\nu}_l)$
by the BaBar collaboration which deviates by 3.4~$\sigma$ (when $D$ and $D^*$ final states are
combined) from the SM prediction~\cite{Lees:2012xj}. If confirmed by BELLE, it means that the
SM is excluded at 3.4~$\sigma$ and also that models type I, II, X and Y will be excluded with a similar
significance. It should be noted that 2HDMs that have the SM as its decoupling limit can never
be excluded in favour of the SM.

\subsection{Production cross sections}
\label{sec:sigma}

In our analysis, we include all production mechanisms of the lightest
scalar $h$ with a mass of 125 GeV.
Processes that involve gauge boson couplings to scalars can be obtained
by simply rescaling the
SM cross section by an appropriate trigonometric factor. Therefore, for
VBF and associated Higgs production
with either a $W$, a $Z$ or with $t\bar{t}$ we have used the results of~\cite{LHCHiggs}.
We have included QCD corrections but
not the SM electroweak corrections
because they can be quite different for the 2HDMs. Cross sections for
$b \bar{b} ~\to h$ are included at
NNLO~\cite{Harlander:2003ai}
and the gluon fusion process $gg \to h$ was calculated using
HIGLU~\cite{Spira:1995mt}.
There are a number of chain processes that lead to final states
with at least one light Higgs~\cite{Djouadi:1999rca}.
From those
we have singled out the ones that give the largest contribution to the
number of light Higgs in the final state.
As a rule, all production processes have at least one of the scalars
$H$, $A$ or $H^\pm $ as an intermediate state that
subsequently decays to a final state that includes at least one
125 GeV Higgs boson. For the gluon
fusion and $b \bar{b}$ initiated processes in $pp \to S$, where $S=H,A$
we again use bb@nnlo and HIGLU.
Note that the pseudo-scalar production has a much more pronounced peak
at the $t \bar{t}$ threshold than the corresponding
CP-even production because the $t \bar{t}$ bound state is also CP-odd.
Regarding the heavy CP-even
scalar associated production we again use~\cite{LHCHiggs}. The non-resonant production of
$gg (b \bar b) \to hh$ was
shown to be negligible for most of 2HDM parameter space when compared to
resonant production via a heavy
CP-even Higgs~\cite{Arhrib:2009hc}. The same is true for
$gg (b \bar b) \to hH$.
A charged Higgs with a mass below the top threshold is mainly produced in
top decays, $t \to H^+ \bar b$, originating
from $pp \to t \bar{t}$. However, such a light charged Higgs would not contribute
significatively to the chain decays. Therefore we have considered cross sections for production of
charged Higgs with masses above 200 GeV. Moreover, due to the flavour
constraints on the charged Higgs mass, the type II
charged Higgs mass is taken to be above 360 GeV. For heavy charged Higgs,
the main production process are $pp \to t H^+$ and $pp \to t \bar{b} H^+$,
and the numerical values for this cross section were obtained
from~\cite{Huitu:2010ad}. The remaining single charged Higgs
production processes were shown to be less important~\cite{Aoki:2011wd}
while both $pp \to H^+ H^-$ and
$pp \to H^\pm W^\mp$ where shown to be small for non-resonant
production~\cite{Aoki:2011wd}. In view of the limits
on the Higgs couplings in 2HDMs~\cite{Ferreira:2013qua} as a consequence
of the discovery of a light Higgs boson,
it is found that we must have a value of $\sin (\beta - \alpha)$ close to $1$.
As such, the process
$qq' \to W^+ \to h H^+$
should be small because it has a coupling proportional to $\cos (\beta - \alpha)$
while $q\bar q' \to (W^+)^* \to H (A) H^+$
are phase space suppressed if $H$ or $A$ are allowed to decay to a final
state with a light Higgs.

\section{Current bounds on heavy states in the 2HDM}
\label{sec:heav}

Since there are considerable uncertainties in the hadronic cross sections for
$h$ production at the LHC, the best observables to compare theory with
experimental results are ratios between observed rates of events {\em versus}
what was expected to be observed in the case of a SM Higgs. Namely, if one is
interested in the production of a lightest Higgs scalar $h$ being produced and decaying
into a given final state $f$
(currently, $f$ = WW, ZZ, $\gamma\gamma$, $b\bar b$ or $\tau \bar \tau$)
in the context of the 2HDM, the ratio we will be
comparing with data is defined as
\be
R^h_f \,=\, \frac{\sigma^{2HDM} (pp\rightarrow h) \,BR^{2HDM}(h \rightarrow f)}
{\sigma^{SM} (pp\rightarrow h) \,BR^{SM}(h \rightarrow f)},
\label{eq:Rh}
\ee
where production cross sections $\sigma$ include all possible {\em direct} production mechanisms
(see section~\ref{sec:sigma}), both for the SM and 2HDM. Likewise, the
branching ratios (BR) of the lightest Higgs boson are computed for both models. Notice
the superscript ``$h$", which will become important later.

In a similar manner, one defines ratios pertaining to the searches for the heavy CP-even
scalar $H$ and pseudoscalar $A$. The ratios take as a reference the production of a SM-like
scalar with mass identical to $H$ or $A$:
\be
R^H_f = \frac{\sigma^{2HDM} (pp\rightarrow H) \,BR^{2HDM}(H \rightarrow f)}
{\sigma^{SM} (pp\rightarrow H) \,BR^{SM}(H \rightarrow f)} \;\;\;,\;\;\;
R^A_f = \frac{\sigma^{2HDM} (pp\rightarrow A) \,BR^{2HDM}(A \rightarrow f)}
{\sigma^{SM} (pp\rightarrow H) \,BR^{SM}(H \rightarrow f)}\,\,.
\ee

\begin{figure}[htb]
\centering
{\epsfysize=7cm
\epsfbox{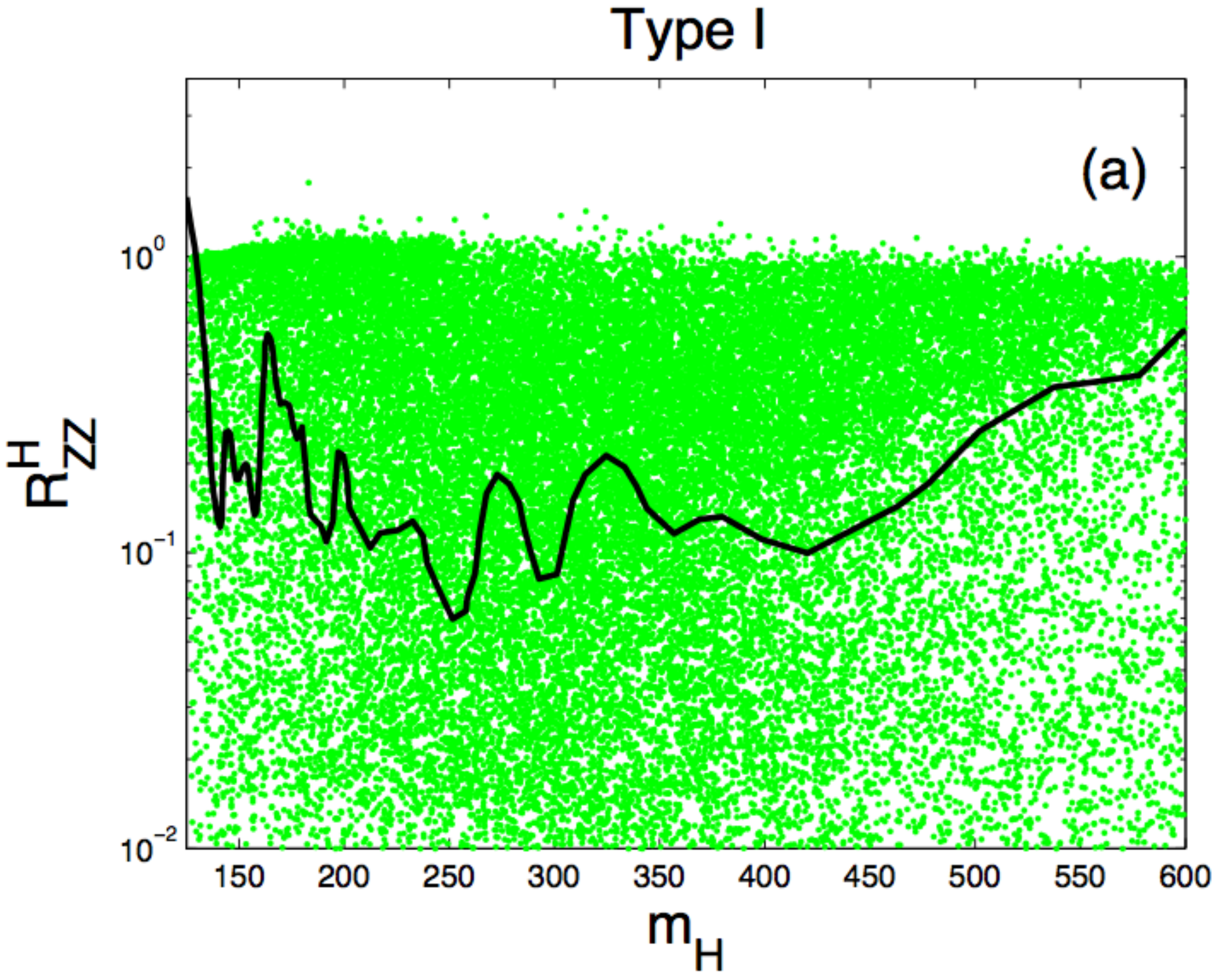}}
{\epsfysize=7cm
\epsfbox{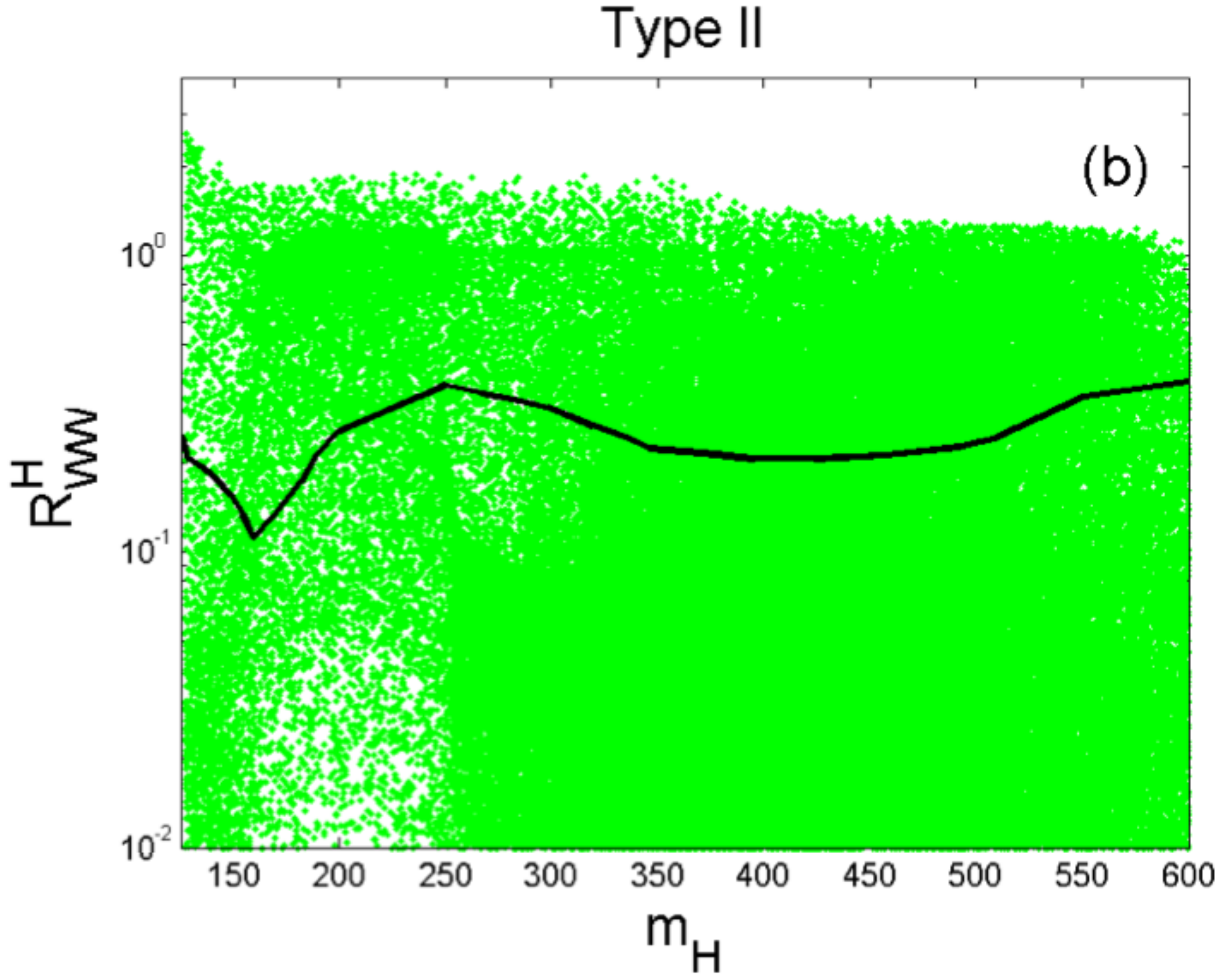}}
\caption{Comparison between 2HDM predictions and LHC results. (a), the
predicted values for $R^H_{ZZ}$, for model type-I. (b), $R^H_{WW}$, for model type-II.
All points include perturbative unitarity, stability,
electroweak precision and B-physics cuts. The black line
is the experimental exclusion line from LHC.}
\label{fig:vv}
\end{figure}

In fig.~\ref{fig:vv} (a) we see the results of our scan of 2HDM parameter space, for
model type-I, for the observable $R^H_{ZZ}$. The black line corresponds to the current experimental
bound for this variable~\cite{atlas_zzww,cms_zzww} - all points below this line are still allowed, all points
above it are excluded. Green (grey) points shown in this figure include perturbative unitarity,  stability,
electroweak precision and B-physics cuts.

As we see from fig.~\ref{fig:vv} (a), the current experimental bounds on $R^H_{ZZ}$ already
exclude a large portion of possible 2HDM points - however, notice that there is still a great amount
of 2HDM parameter space allowed, for all the range of masses $m_H$ considered. Analogous results
are obtained for $R^H_{ZZ}$ in model type-II.

In a similar manner, LHC data on $R^H_{WW}$~\cite{atlas_zzww,cms_zzww} exclude many possible combinations
of 2HDM parameters,
while still allowing many others. We show this for model type-II (analogous results
hold for model type-I) in fig.~\ref{fig:vv} (b). Please notice that for type-II the parameter
space scan already included the B-physics (and LHC) motivated restriction, $m_{H^\pm} > 360$ GeV~\cite{BB2}.

What these plots are telling us is quite simple: the heaviest CP-even scalar $H$, if it exists, should
not couple too strongly to gauge bosons. This is hardly surprising, given that the LHC
data indicates that the lightest CP-even state, $h$, seems to couple to gauge bosons with
practically its SM-expected strength. Now, gauge symmetry implies the following
sum rule for the couplings of $h$ and $H$ to the gauge bosons:
\be
g^2_{hVV} \,+\,g^2_{HVV}\,=\,1\, .
\ee
If $h$ is almost SM-like then $g_{hVV} \simeq 1$, which means $g_{HVV} \simeq 0$ - thus it is to
be expected that $H$ couples weakly to gauge bosons, making it more difficult to observe that scalar
state in this channel.

Some care must be taken while comparing the $ZZ$ and $WW$ calculated rates at such high values
of $m_H$. In fact, there is the possibility that, if the scalar's width is too large, the interference
of the process $pp \rightarrow h \rightarrow ZZ/WW$ with the $ZZ/WW$ background becoming relevant.
Previous studies~\cite{zzback} analysed this possibility, which was shown to, at most, provoke a
reduction of 10\% on the observed signal. To be conservative, then, we will ``add" 10\% to the
experimental exclusion line of figs.~\ref{fig:vv}, left and right. This way we accept all 2HDM points
which lie below that line or which, at most, live above 10\% of its value for a given choice of
$m_H$.

\begin{figure}[htb]
\centering
{\epsfysize=7cm
\epsfbox{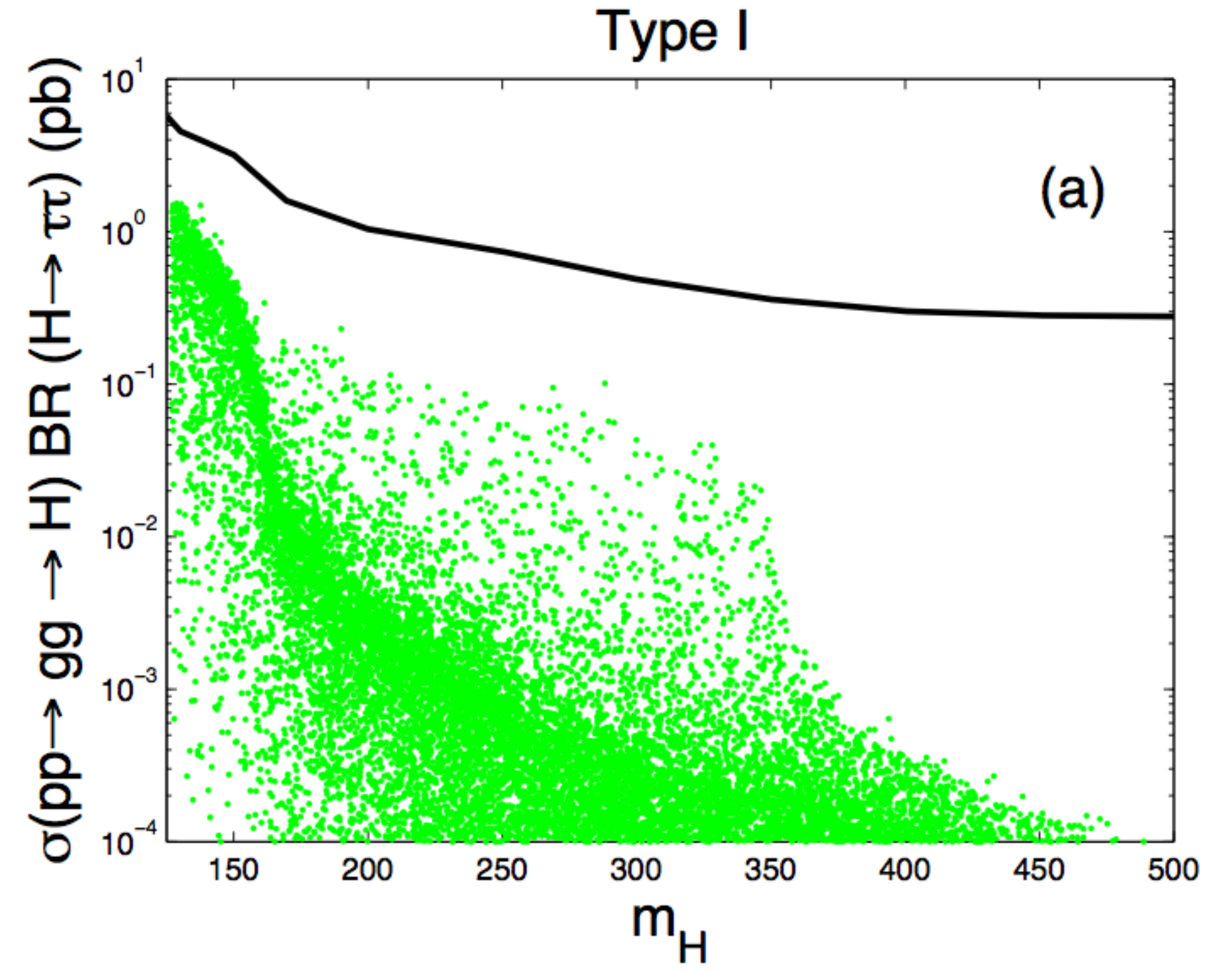}}
{\epsfysize=7cm
\epsfbox{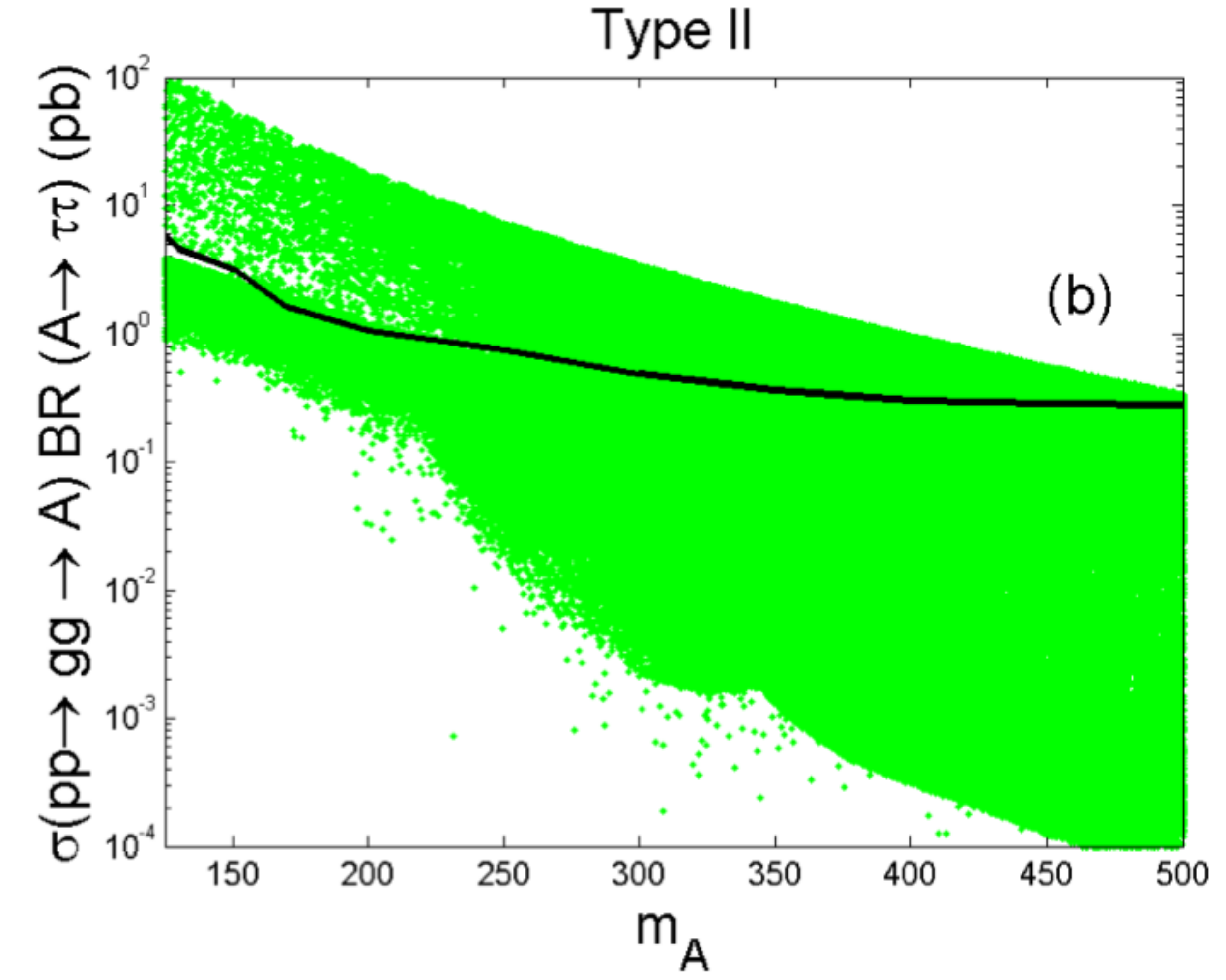}}
\caption{Comparison between 2HDM predictions and LHC results: (a), the
predicted values for $\sigma(pp\rightarrow H) \times BR(H\rightarrow \tau\tau)$, for model type-I.
(b), $\sigma(pp\rightarrow A) \times BR(A\rightarrow \tau\tau)$, for model type-II.
All points include perturbative unitarity, stability,
electroweak precision and B-physics cuts. The black line
is the experimental exclusion line from LHC.}
\label{fig:tau}
\end{figure}
In fig.~\ref{fig:tau} (a) we show the results of our parameter space scan for the product
$\sigma(pp\rightarrow H) \times BR(H\rightarrow \tau\bar{\tau})$ for model type-I, and compare it to
the current bound from LHC~\footnote{Unlike all other channels, for the $\tau\bar{\tau}$ results at
high masses the LHC collaborations have published results for $\sigma \times BR$.}~\cite{atlas_tautau}.
As we see, there is still no exclusion coming from the
$\tau\bar{\tau}$ channel (which is
understandable, considering that even for the $h$ scalar the knowledge of this channels is still
in its infancy). For model type-II, however, we would find a sizeable portion of parameter space
{\em above} the LHC exclusion line, so the $\tau\bar{\tau}$ data is already helping us putting
constraints on the 2HDM. The reason why one obtains a more pronounced exclusion in model type-II
is the fact that the Higgs coupling to the bottom quarks, and to the tau leptons, increases significantly with
$\tan\beta$ (something which does not occur in model type-I). Thus both the production and decays of
$h$ are enhanced for large values of $\tan\beta$.

Likewise, the $\tau\bar{\tau}$ exclusion is already extremely useful in constraining the existence of the
pseudoscalar $A$. As we see from fig.~\ref{fig:tau} (b), the same bound we used in the production and decay
of $H$ already excludes a sizeable
region of parameter space due to the decays of $A$ to $\tau\bar{\tau}$ in model type-II. The exclusion
is less pronounced for model type-I.
The reason for the added importance of the $\tau\bar{\tau}$ channel in the
pseudoscalar's decays is simple to understand: for $A$ masses below $\sim$350 GeV (the threshold for
$A\rightarrow t\bar{t}$ decays), the decays of $A$ into $b\bar{b}$ and $\tau\bar{\tau}$ are dominant
over all others~\footnote{Certainly dominant over loop decays such as $A\rightarrow \gamma\gamma$ and
$A\rightarrow Z\gamma$, and favoured over the decay $A\rightarrow Zh$ (which is proportional to $\cos(\beta - \alpha)$,
even above its threshold of $\sim$216 GeV, for many choices of parameters in the model.}. This dominance clearly does not
occur for the CP-even state $H$, which decays predominantly to WW or ZZ in the same mass range. As such,
the expected $\sigma \times BR$ for $\tau$'s are larger for $A$ than for $H$, explaining the difference in
behaviour in figs.~\ref{fig:tau} (a) and~\ref{fig:tau} (b). The excluded region has mostly high values of $\tan\beta$
since they enhance immensely the production cross section in model type-II~\cite{tau}, due to the coupling
of the pseudoscalar to the bottom quarks being proportional to $\tan\beta$~\footnote{For those high values of
$\tan\beta$ the pseudoscalar coupling to the top quark is suppressed, which explains why we do not see a decrease on
$\sigma(pp\rightarrow A) \times BR(A\rightarrow \tau\tau)$ in fig.~\ref{fig:tau} (b) around $2 m_t$.}.

\begin{figure}[htb]
\centering
{\epsfysize=7cm
\epsfbox{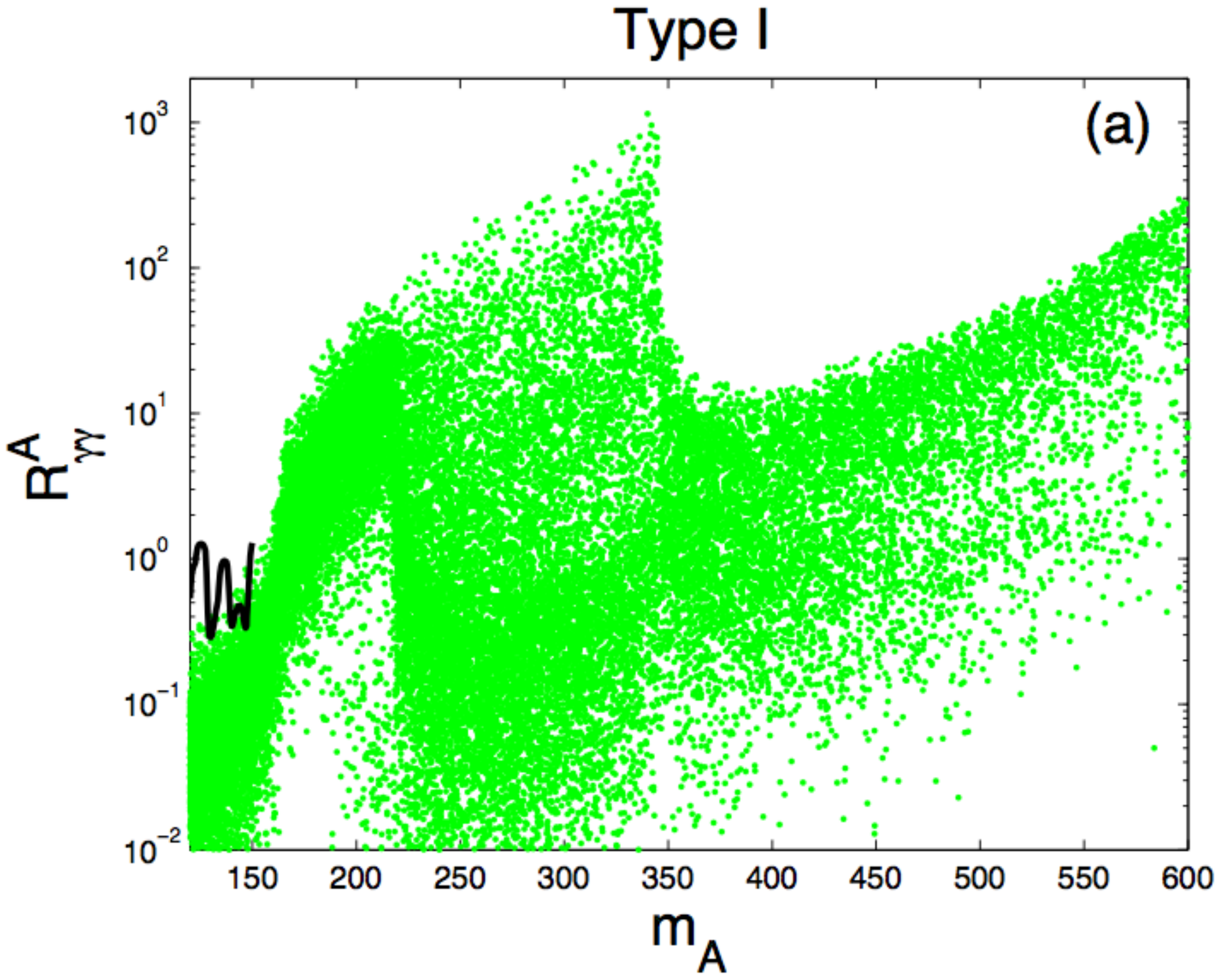}}
{\epsfysize=7cm
\epsfbox{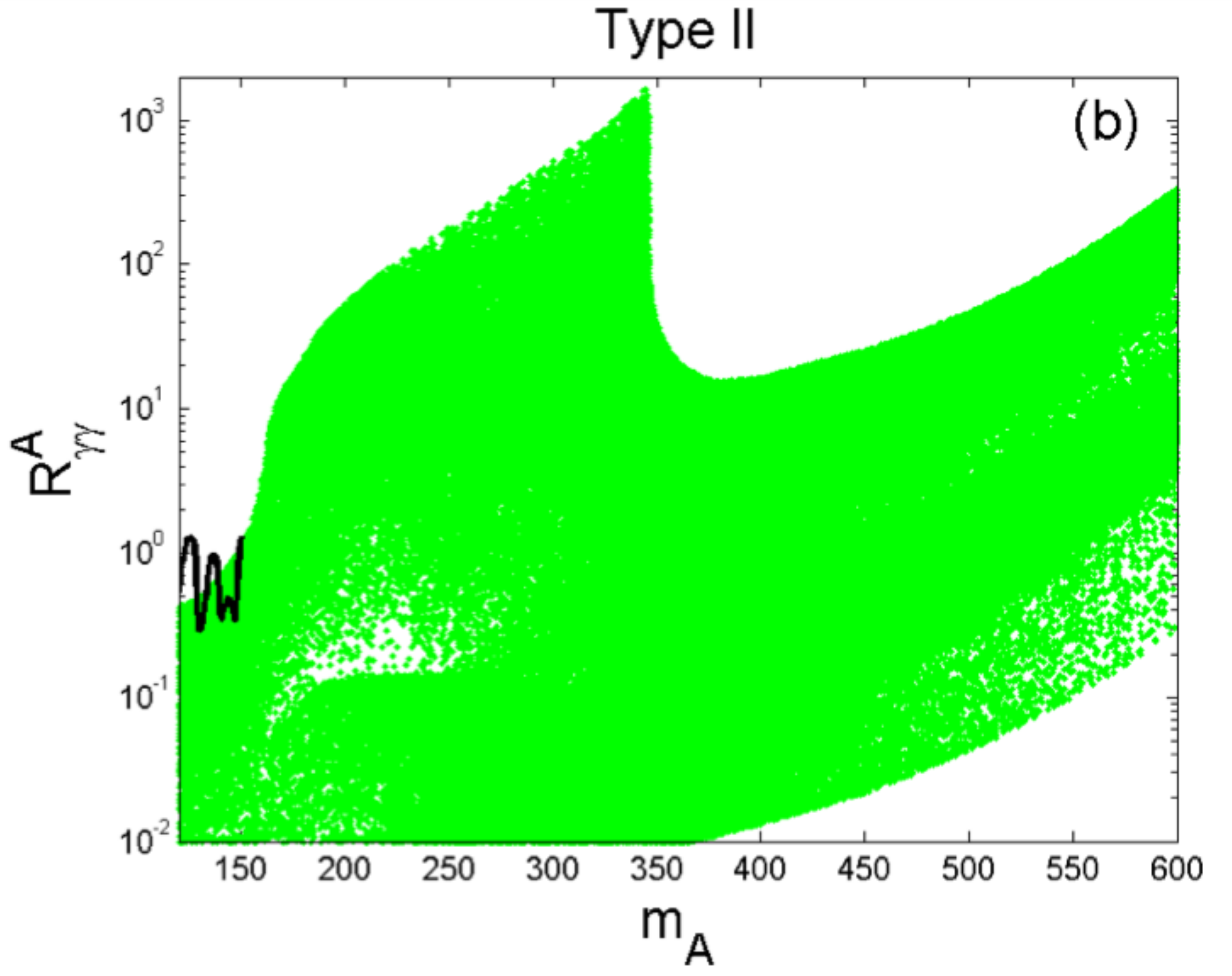}}
\caption{Comparison between 2HDM predictions and LHC results. (a), the
predicted values for $R^A_{\gamma\gamma}$, for model type-I.
(b), $R^A_{\gamma\gamma}$, for model type-II.
All points include perturbative unitarity, stability,
electroweak precision and B-physics cuts. The black line
is the experimental exclusion line from LHC.}
\label{fig:gamma}
\end{figure}
Finally, the $\gamma\gamma$ data can also be used to search for a second peak which might reveal the
presence of the $H$ and/or $A$. The LHC collaborations provides us with an exclusion curve for
$R_{\gamma\gamma}$ for several masses~\cite{atlas_phph,cms_phph}, but the published results
go only up to masses of the pseudoscalar of 150 GeV. In
fig.~\ref{fig:gamma} we see the results of our scans for models I and II, for the pseudoscalar
rate $R^A_{\gamma\gamma}$ (similar results hold for $R^H_{\gamma\gamma}$ in both models). Like for the
$\tau\bar{\tau}$ channel, not much of the parameter space is excluded. However,
it is worth pointing out that it is possible to obtain, in the 2HDM, extremely large enhancements
of the diphoton signal. The sharp decrease we observe around masses of 350 GeV is clearly due to the
opening of the decay channel $A \rightarrow t\bar{t}$, but before that we can have enhancements of
a factor of 1000 or more over the expected SM value. Now, the diphoton channel is very difficult to
analyse, but we would expect that such large enhancements in the signal coming from $A$ can easily be
excluded by existing data. It would be interesting to obtain experimental exclusion curves in the
high mass region for the diphoton channel as well, since it is clear that a great deal of 2HDM parameter
space may well be ruled out by that analysis.

\begin{figure}[htb]
\centering
{\epsfysize=7cm
\epsfbox{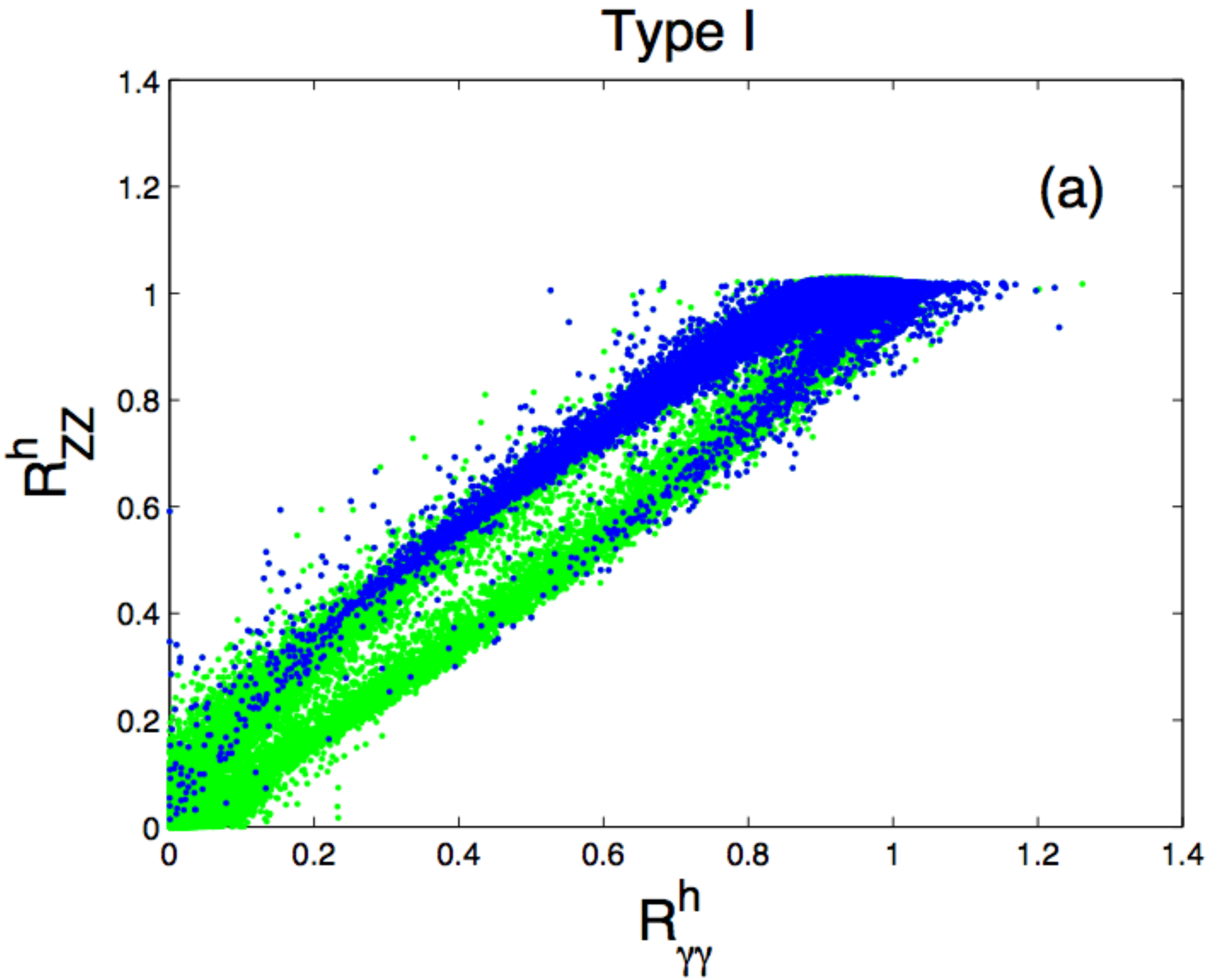}}
{\epsfysize=7cm
\epsfbox{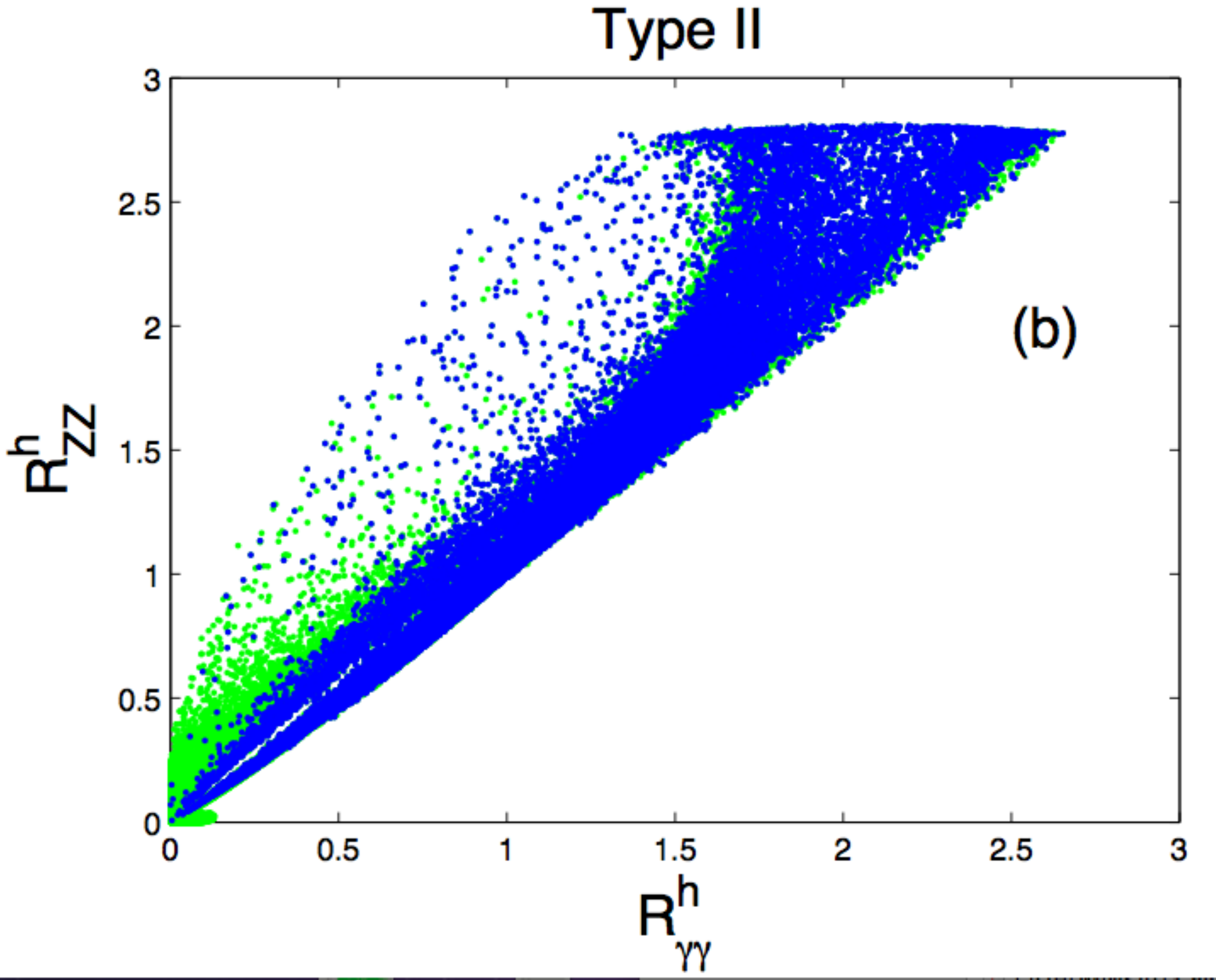}}
\caption{Predictions, within the 2HDM, for $R^h_{ZZ}$ {\em vs} $R^h_{\gamma\gamma}$,
for model type-I (a) and type II (b). Green (grey) points include perturbative unitarity, stability,
electroweak precision and B-physics cuts.
Blue (black) points include, on top of that, the exclusion of 2HDM parameters from
the non-observation of $H$ and $A$.}
\label{fig:hpZ}
\end{figure}
In short, the current LHC results allow us to exclude portions of 2HDM parameter space, from the
non-observation of a second peak corresponding to $H\rightarrow WW$ or $H\rightarrow ZZ$ for masses
$125 \lesssim m_H \lesssim 600$ GeV. The $\tau\bar{\tau}$ data provide us with nice restrictions due to the
non-observation of a pseudoscalar $A$ and, in model type-II, also from signals coming from $H$. The
$\gamma\gamma$ channels also give some
mild exclusion, again mostly due to the pseudoscalar $A$. But if we combine all of these exclusions,
we conclude there is still a great deal of 2HDM parameter space allowed, with extra scalar masses
in ranges from 125 to 600 GeV (or even larger) completely allowed. Neither does the
excluded region have any dramatic consequences for the lightest $h$ observables. As we see from
figs.~\ref{fig:hpZ} (a) and (b), the non-observation of $H$ and $A$ does give some differences in the
allowed predictions for $R^h_{\gamma\gamma}$ and $R^h_{ZZ}$. But since the current best values for these
observables are, respectively, $1.55 \tiny{\begin{array}{l}+0.33\\ -0.26\end{array}}$ and $1.5 \pm 0.4$ (ATLAS) or
$0.78 \tiny{\begin{array}{l}+0.28\\ -0.26\end{array}}$ and $0.91 \pm 0.30$ (CMS), we conclude
that there is nothing dramatic one can thus far conclude from the non-observation of neutral heavy states
at the LHC. However, future improvements on the $\tau\bar{\tau}$ exclusion bounds for high masses,
and the extension of the $\gamma\gamma$ exclusion to the high mass regime, hold the potential for
severe restrictions on the 2HDM parameter space.

It is interesting to profile the pseudo-scalars that can give the largest contribution to chain
decays. Their mass should be above the $hZ$ threshold and not too large. The values of $\tan\beta$
should close to 1 in both models. Moreover, we have checked that the preferred values of $\sin(\beta - \alpha)$
are those close to 1. Because $\Gamma(A\rightarrow hZ) \propto \cos(\beta - \alpha)$, and
$\cos(\beta - \alpha)\sim 0$, this width is small. Hence, to obtain a significant $BR(A\rightarrow hZ)$,
the total width has to be small as well (below 1 GeV, at least). Further, the low values of $\tan\beta$
enhance the production cross sections of the pseudoscalar, via gluon-gluon fusion.

\section{Higgs chain production}
\label{sec:chain}

If the heaviest CP-even state $H$ is heavy enough, it can decay to the lighter $h$
via several possibilities - for instance, the decay channel $H\,\rightarrow \,hh$.
Since the production cross sections for $H$, for certain
2HDM parameter choices, can actually be quite large, we see that there is another
possibility of lightest Higgs production at the LHC: the heavy state $H$ is produced
and then decays to $h$. We call this process Higgs chain production, and $H$ is not the
single contributor to it: the pseudoscalar $A$ and the charged Higgs $H^\pm$ can also
have decays in which $h$ is present, and as such may contribute to the number of lightest
Higgs produced at the LHC.

In fact, chain production raises an interesting possibility: the decays of the heavier states
to the lightest $h$ may correspond to the {\em dominant} branching ratio of those particles.
For instance, when the decay channel $H\,\rightarrow \,hh$ is open, for a wide range of
values of 2HDM parameters this decay is dominant over all others. It is easy to understand
why: since the Higgs scalars couple proportionally to the mass, for an $H$ mass of, say,
300 GeV, decaying to two h's is favoured over decaying to two W's or two Z's, or any fermions -
unless the triple coupling $Hhh$ is accidentally small, which is possible with some parameter
fine-tuning. As such, for those (ample) regions of parameter space $H$ could not be detected
via its decays to fermions or gauge bosons, and only the study of $h$ production could
infer the presence of the heavier CP-even state. Similar circumstances, involving other channels,
can also occur for the pseudoscalar $A$ or the charged scalar $H^\pm$.
\begin{figure}[htb]
\centering
{\epsfysize=7cm
\epsfbox{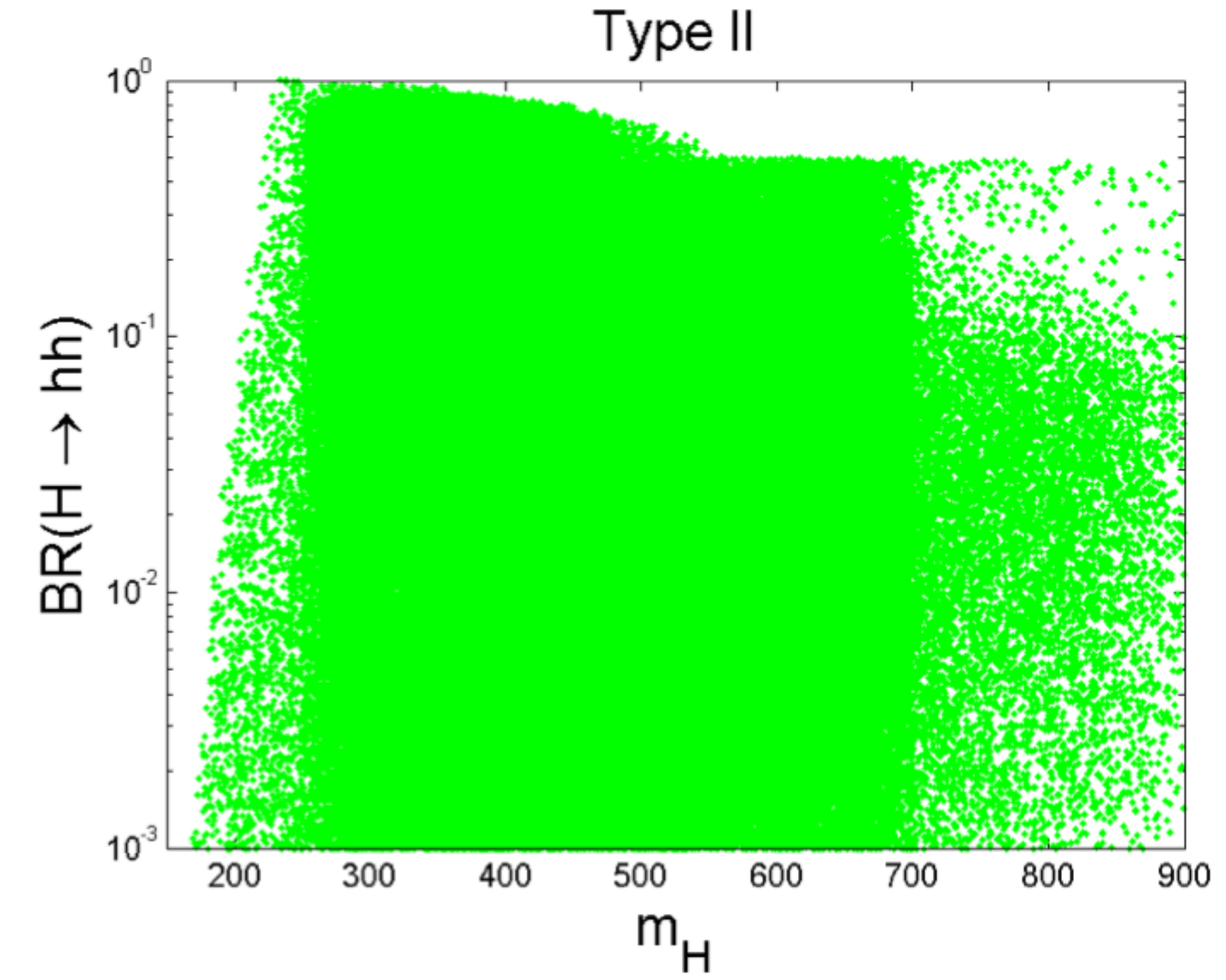}}
{\epsfysize=7cm
\epsfbox{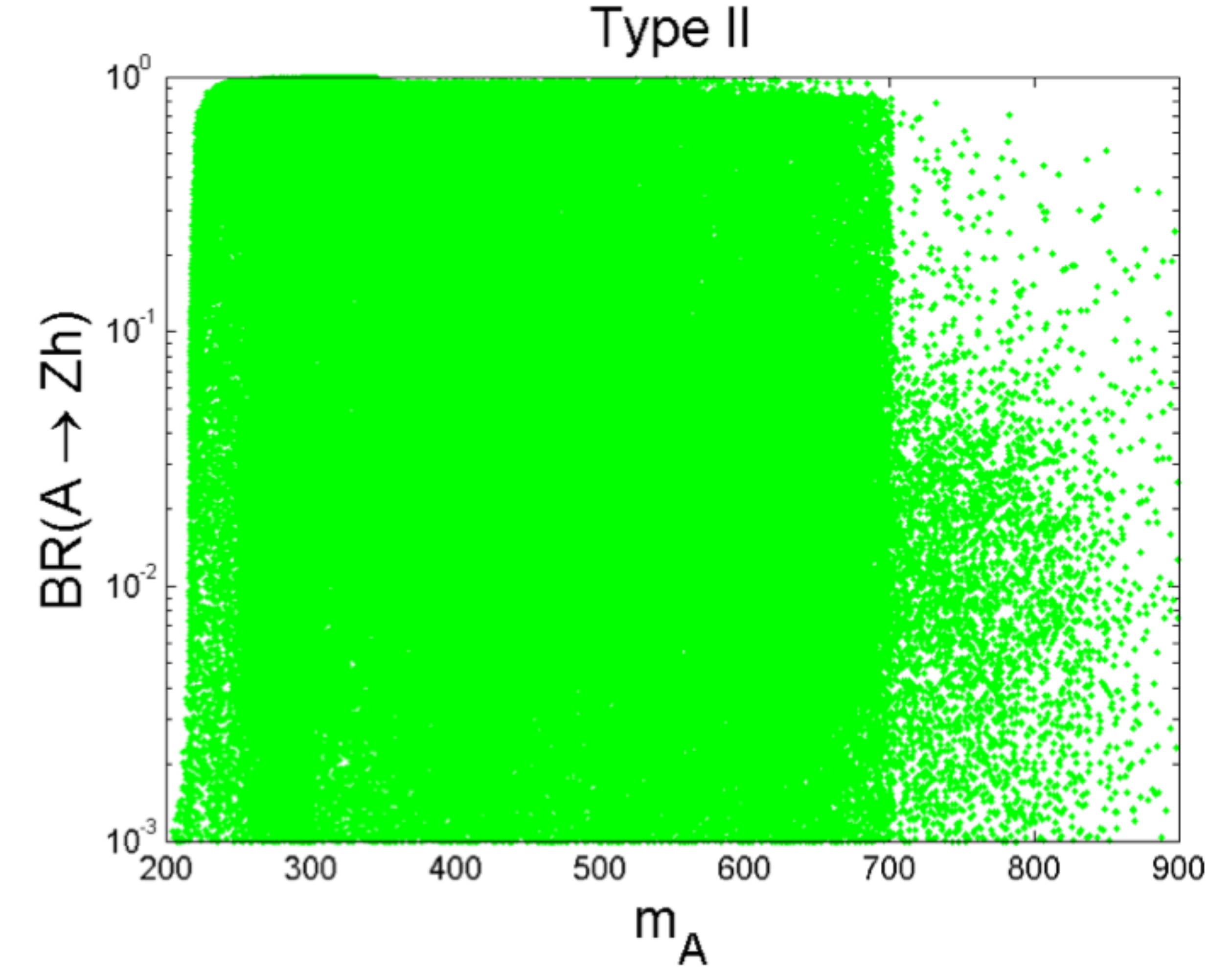}}
\caption{Branching ratios for the decays $H\,\rightarrow \,hh$ and $A\,\rightarrow \,Zh$
in terms of $m_H$ and $m_A$ for model type-II.
All non-LHC (B-physics, perturbative unitarity, stability and electroweak precision) constraints have been
imposed. Analogous results hold for model type-I.}
\label{fig:brs}
\end{figure}
Chain Higgs production is thus an alternate source of LHC lightest Higgs scalars, and we wish to
verify whether it can have any impact on current observables. As we see from fig.~\ref{fig:brs},
the branching ratios for the decays $H\,\rightarrow \,hh$ and $A\,\rightarrow \,Zh$ can indeed
be quite large and close to one, these becoming the dominant decays for a wide region of
parameter space. Likewise for the decay $H\pm \rightarrow W^\pm h$.

\subsection{New contributions to observable rates}

The rates shown in eq.~\eqref{eq:Rh}
are {\em partial} ones, since they do not take into account $h$ indirect production through
the decays of the heavier scalars. As such, the observed/expected {\em total} ratios for the lightest
scalar $h$ will have two contributions:
\be
R^{T,h}_f \,=\, R^h_f \,+\, R^C_f\,
\label{eq:nrat}
\ee
where we see the contribution of eq.~\eqref{eq:Rh} which describes direct $h$ production;
and a new contribution from chain $h$ production, {\em a priori} including contributions from
all the other scalar states (when suitably heavy):
\bea
R^C_f &=& R^{C,H}_f \,+\, R^{C,A}_f \,+\, R^{C,H^\pm}_f.
\eea
These contributions are given by, for $H$,
\be
R^{C,H}_f \,=\, \frac{\sigma^{2HDM} (pp\rightarrow H)}{\sigma^{SM} (pp\rightarrow h)} \,
N_{H,h}\, \frac{BR^{2HDM}(h \rightarrow f)}{BR^{SM}(h \rightarrow f)},
\label{eq:RCH}
\ee
where $N_{H,h}$ is the expectation value of the number of lightest Higgs scalars $h$
produced in the decays of $H$, {\em i.e.}
\be
N_{H,h} = 2\times P_{H,2h} \,+\, 1\times P_{H,1h}
\ee
where $P_{H,2h}$ is the probability of finding two $h$ scalars in decays of $H$, and
$P_{H,1h}$ the probability of obtaining a single $h$. Taking into account all
possible decays, we will have~\footnote{We thank J. Yun for useful discussions concerning these formulae.}
\begin{eqnarray}
P_{H,2h} & = & BR(H \rightarrow hh) + \nonumber \\
 &  & BR(H \rightarrow H^+ H^-)\,\left[ BR(H^+ \rightarrow W^+ h)^2 \,+\, BR(H^+ \rightarrow W^+ A)^2
BR(A \rightarrow Z h)^2 \, \right. + \nonumber \\
 &  &  \quad\quad\quad\quad\quad\quad\quad\quad\quad
 \left. 2 BR(H^+ \rightarrow W^+ h)\, BR(H^+ \rightarrow W^+ A)\,BR(A \rightarrow Z h) \right] \,+ \nonumber \\
  & & BR(H \rightarrow AA)\,\left[BR(A \rightarrow Z h)^2 \,+\, 4 BR(A \rightarrow W^- H^+)^2
BR(H^+ \rightarrow W^+ h)^2 \, + \right. \nonumber \\
 &  &  \quad\quad\quad\quad\quad\quad\quad\quad\quad
 \left. 4 BR(A \rightarrow Z h)\, BR(A \rightarrow W^- H^+)\,BR(H^+ \rightarrow W^+ h) \right]
\end{eqnarray}
as well as
\begin{eqnarray}
P_{H,1h} &=& BR(H \rightarrow Z A) \,BR(A \rightarrow Z h) \,+\, 2 BR(H \rightarrow W^- H^+ ) \,BR(H^+ \rightarrow W^+ h) \,+\,
\nonumber \\
  & & 2 BR(H \rightarrow AA)\,\left[ BR(A \rightarrow Z h) \, + \,
  2 BR(A \rightarrow W^- H^+)\,BR(H^+ \rightarrow W^+ h)\right]\,p(A\slashed{\rightarrow}h)\, + \nonumber \\
  & & 2 BR(H \rightarrow H^+ H^-)\,\left[ BR(H^+ \rightarrow W^+ h) \,+\,
  BR(H^+ \rightarrow W^+ A)\,BR(A \rightarrow Z h)\right]\,p(H^+\slashed{\rightarrow}h).
\end{eqnarray}
Finally, we have to define the probabilities that $A$ or $H^+$ do {\em not} decay into $h$,
given by
\begin{eqnarray}
p(A\slashed{\rightarrow}h) &=& 1 \,-\, BR(A \rightarrow Z h) \,-\,
2 BR(A \rightarrow W^- H^+)\,BR(H^+ \rightarrow W^+ h)\, , \nonumber \\
p(H^+\slashed{\rightarrow}h) &=& 1 \,-\, BR(H^+ \rightarrow W^+ h)\,-\,
BR(H^+ \rightarrow W^+ A)\,BR(A \rightarrow Z h)\,.
\end{eqnarray}
Please notice that many of the decays present in these formulae may well be
kinematically forbidden for many choices of 2HDM parameters, in which case the corresponding
branching ratios will automatically be zero~\footnote{Notice, however, that like for the $WW$
and $ZZ$ channels, we also utilized off-shell formulae for the decays of heavier scalars into 
$h$~\cite{DKZ}.}. 
So, for instance, the decay $H \rightarrow hh$ will only be
relevant if $m_H > 2 m_h \simeq 250$ GeV, the decay $A \rightarrow Z h$ requires $m_A > m_Z + m_h
\simeq 216$ GeV, and so on.

Likewise, we will have, for $h$ chain production via pseudoscalar $A$ production, the
following contribution to the rate ratios:
\be
R^{C,A}_f \,=\, \frac{\sigma^{2HDM} (pp\rightarrow A)}{\sigma^{SM} (pp\rightarrow h)} \,
N_{A,h}\, \frac{BR^{2HDM}(h \rightarrow f)}{BR^{SM}(h \rightarrow f)},
\label{eq:RCA}
\ee
where the expectation value of the numbers of $h$ scalars produced through decays of $A$
is given by
\be
N_{A,h} = 2\times P_{A,2h} \,+\, 1\times P_{A,1h}
\ee
where the decay probabilities are
\begin{eqnarray}
P_{A,2h} &=& BR(A \rightarrow Z H) \,\left[ BR(H \rightarrow hh) \,+\,
BR(H \rightarrow H^- H^+ ) \,BR(H^+ \rightarrow W^+ h)^2 \right]\, + \nonumber \\
 & & 2 BR(A \rightarrow W^- H^+)\,BR(H^+ \rightarrow W^+ H) \, BR(H\rightarrow hh), \nonumber \\
P_{A,1h} &=& BR(A \rightarrow Z h) \, + \, 2 BR(A \rightarrow W^- H^+)\,BR(H^+ \rightarrow W^+ h)\, +
\nonumber \\
 & & 2 BR(A \rightarrow Z H) \,\left\{ BR(H \rightarrow H^- H^+)\,BR(H^+ \rightarrow W^+ h)\,
 \left[1 - BR(H^+ \rightarrow W^+ h)\right] \right. \,+\nonumber \\
 & & \quad\quad\quad\quad\quad\quad\quad
 \left. BR(H \rightarrow W^- H^+)\,BR(H^+ \rightarrow W^+ h) \right\}.
\end{eqnarray}
Finally, for the $H^\pm$ contribution, we have
\be
R^{C,H^\pm}_f \,=\, \frac{\sigma^{2HDM} (pp\rightarrow H^\pm)}{\sigma^{SM} (pp\rightarrow h)} \,
N_{H^\pm ,h}\, \frac{BR^{2HDM}(h \rightarrow f)}{BR^{SM}(h \rightarrow f)},
\label{eq:RCHch}
\ee
with
\be
N_{H^\pm ,h} = 2\times P_{H^\pm,2h} \,+\, 1\times P_{H^\pm ,1h}
\ee
where the decay probabilities are
\begin{eqnarray}
P_{H^\pm ,2h} &=& BR(H^+ \rightarrow W^+ H) \,\left[ BR(H \rightarrow hh) \,+\,
BR(H \rightarrow AA ) \,BR(A \rightarrow Z h)^2 \right]\, , \nonumber \\
P_{H^\pm ,1h} &=& BR(H^+ \rightarrow W^+ h) \, + \, BR(H^+ \rightarrow W^+ A)\,BR(A \rightarrow Z h)\, + \nonumber \\
 & & BR(H^+ \rightarrow W^+ H) \,BR(H \rightarrow Z A)\,BR(A \rightarrow Z h)\, .
 \label{eq:nratu}
\end{eqnarray}

\subsection{Chain Higgs production at the LHC}

\begin{figure}[htb]
\centering
{\epsfysize=7cm
\epsfbox{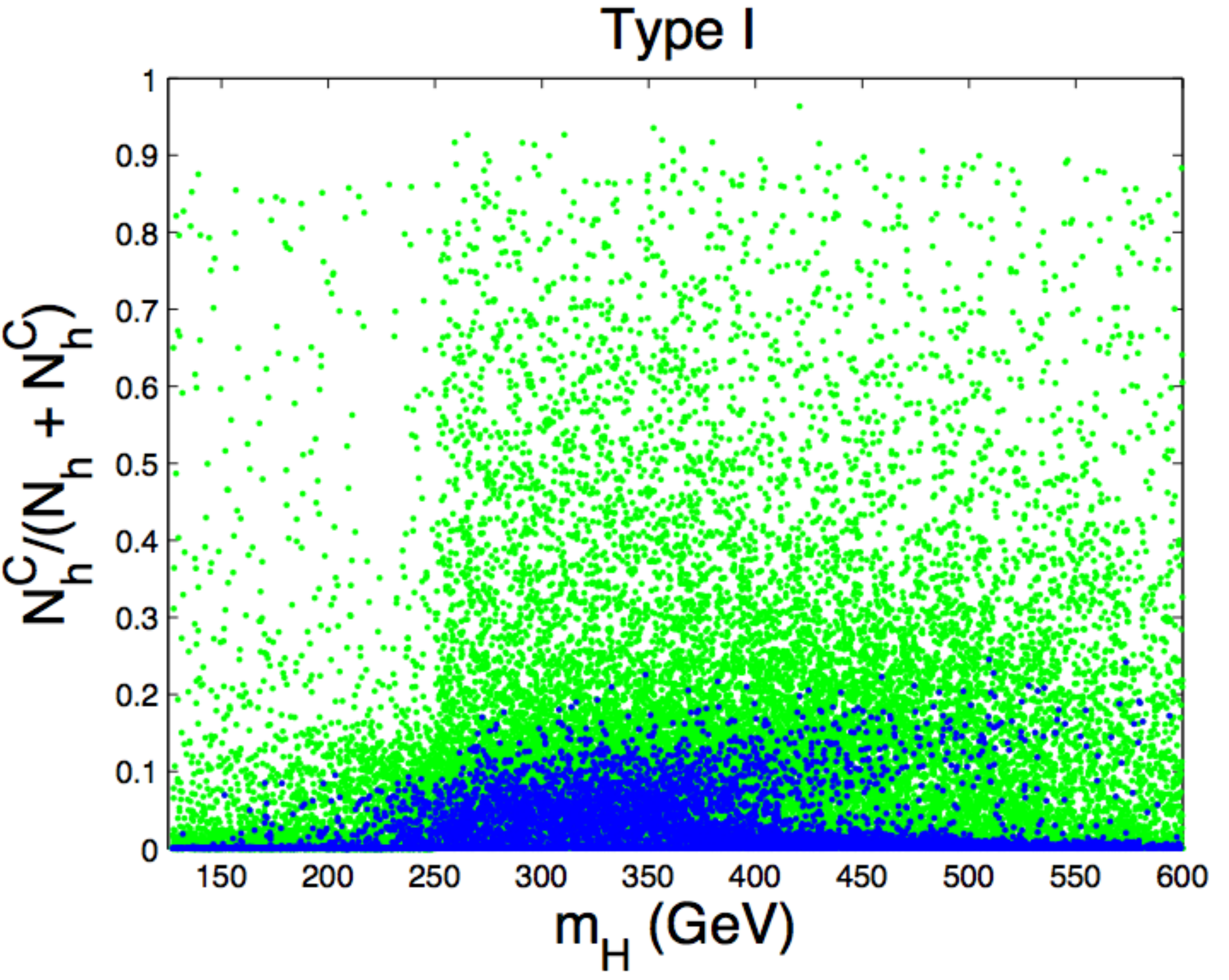}}
{\epsfysize=7cm
\epsfbox{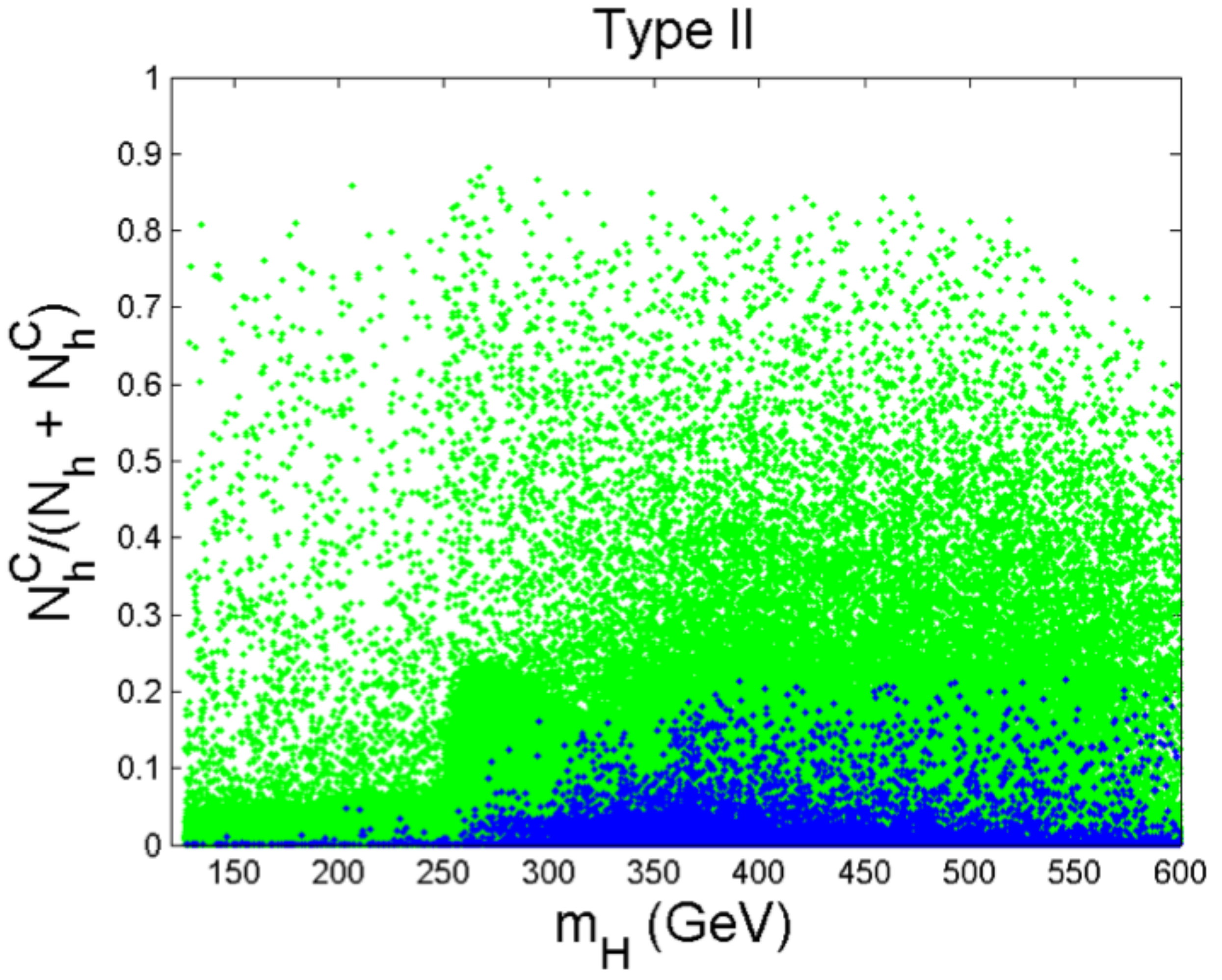}}
\caption{Ratio of number of $h$ scalars coming from chain production to the
total number produced (direct production and chain production) for models I and II.
Green (grey) points include perturbative unitarity, stability,
electroweak precision and B-physics cuts.
In blue (black) we further imposed
cuts coming from non-observation of the heavy scalars $H$ and $A$ {\em and} we also
required that the total rates $R^{T,h}_f$ be within 20\% of their SM values, for all final states $f$.}
\label{fig:Nh}
\end{figure}
With these expressions established, we can begin by looking at the extra number of lightest scalars
$h$ which come from Higgs chain production. In fig.~\ref{fig:Nh} we plot the ratio of the number
of $h$'s produced at the LHC via chain production ($N^C_h$) to the {\em total} number of $h$'s produced
(via chain production and direct production,  $N^C_h + N_h$), for both models type I and II. If one only
requires that non-LHC constraints
are satisfied, we see that chain production can even be the
dominant process for production of the lightest scalar $h$. However, if we further require that the $h$'s
being produced have rates to $ZZ$, $WW$, $\gamma\gamma$, etc, all within 20\% of their expected
SM values~\footnote{By this we mean that the total rates $R^{T,h}_f$ of eq.~\eqref{eq:nrat} are all between
0.8 and 1.2, regardless of the final state $f$.
}, we see that the contributions from chain production are substantially reduced. We have also included
the cuts stemming from the non-observation of $H$ and $A$ at the LHC so far, as detailed in
section~\ref{sec:heav}. Those cuts do not have a major impact on chain Higgs production, as was to be expected:
the regions of parameter space where chain Higgs production is expected
to be significative correspond to large branching ratios for the decays $H\rightarrow hh$ and/or $A\rightarrow Zh$.
Correspondingly, for such choices of parameters the decays of $H$ into $WW$, $ZZ$ or fermions would be
highly suppressed, thus it being natural that $H$ would not have been yet observed in those channels.

\begin{figure}[htb]
\centering
{\epsfysize=7cm
\epsfbox{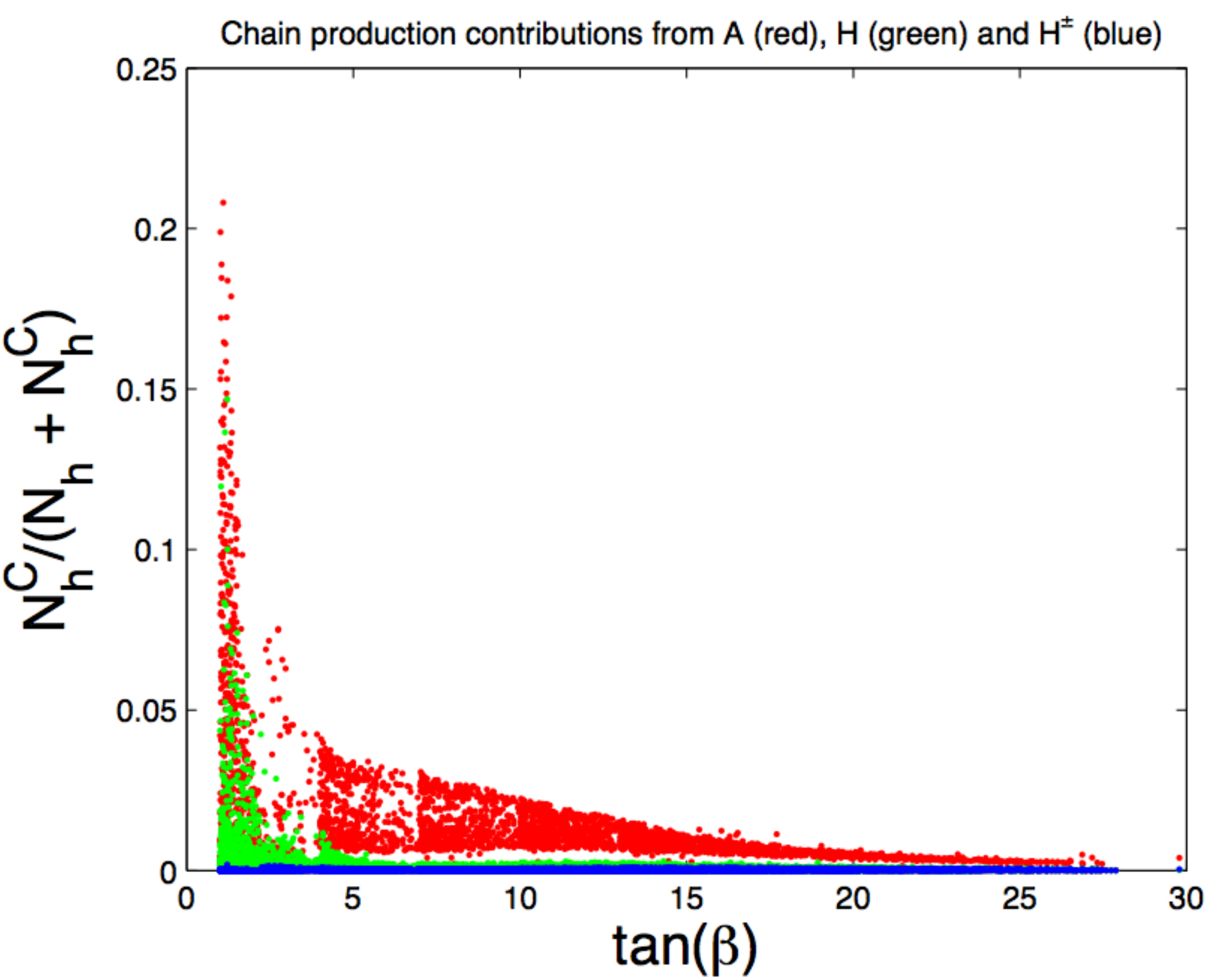}}
\caption{Ratio of number of $h$ scalars coming from chain production to the
total number produced (direct production and chain production) for model type-II.
In red (dark grey) we show the contribution to chain production from the
pseudoscalar $A$; in green (light grey) the contribution from the CP-even heaviest scalar, $H$;
and in blue (black) the contribution from the charged scalar $H^\pm$.
}
\label{fig:Nsvs}
\end{figure}
Still, one sees that chain production can still account for as much as $\sim$ 25\% of the number of
$h$ scalars produced at the LHC - more frequently, chain production can give an extra 10 to 20\%
number of $h$'s at the LHC. Thus, current LHC trends in the observable rates of $h$ do not allow us to
dismiss chain Higgs production as an irrelevant curiosity - it may well have an impact on what we are already
measuring. For completeness, in fig.~\ref{fig:Nsvs} we show the individual contributions to the number of lightest Higgs
coming from chain production from each of the extra scalars~\footnote{Meaning, from the direct production of one of the
extra scalars and subsequent decays to $h$.}, in model type-II
Like before, we demanded that the
$h$ rates be within 20\% of its SM values, as well as all non-LHC cuts.
We see that the charged Higgs $H^\pm$ contribution
to chain production is quite small. The largest contribution stems from the pseudoscalar, via (mostly) the decay
$A\rightarrow Zh$. Slightly smaller, but comparable in size, is the contribution from the heaviest CP-even scalar
$H$, via (mostly) the decay $H\rightarrow hh$. Also, from fig.~\ref{fig:Nsvs}, we ascertain that the largest
contributions to chain production occur for relatively smaller values of $\tan\beta$ ({\em i.e.}, smaller than about
$5$). Finally, the off-shell contributions from decays such as $A\rightarrow Zh$ or $H\rightarrow hh$ are 
much smaller than the on-shell ones, and have a negligible effect on the final results. 

\subsection{Chain Higgs production contributions to observable $h$ rates}

Having established that chain Higgs production may have a sizeable contribution
to the number of $h$ scalars seen at the LHC, we need then to ask how one might
detect this phenomenon. Let us first consider the rates of production and decay
\begin{figure}[htb]
\centering
{\epsfysize=7cm
\epsfbox{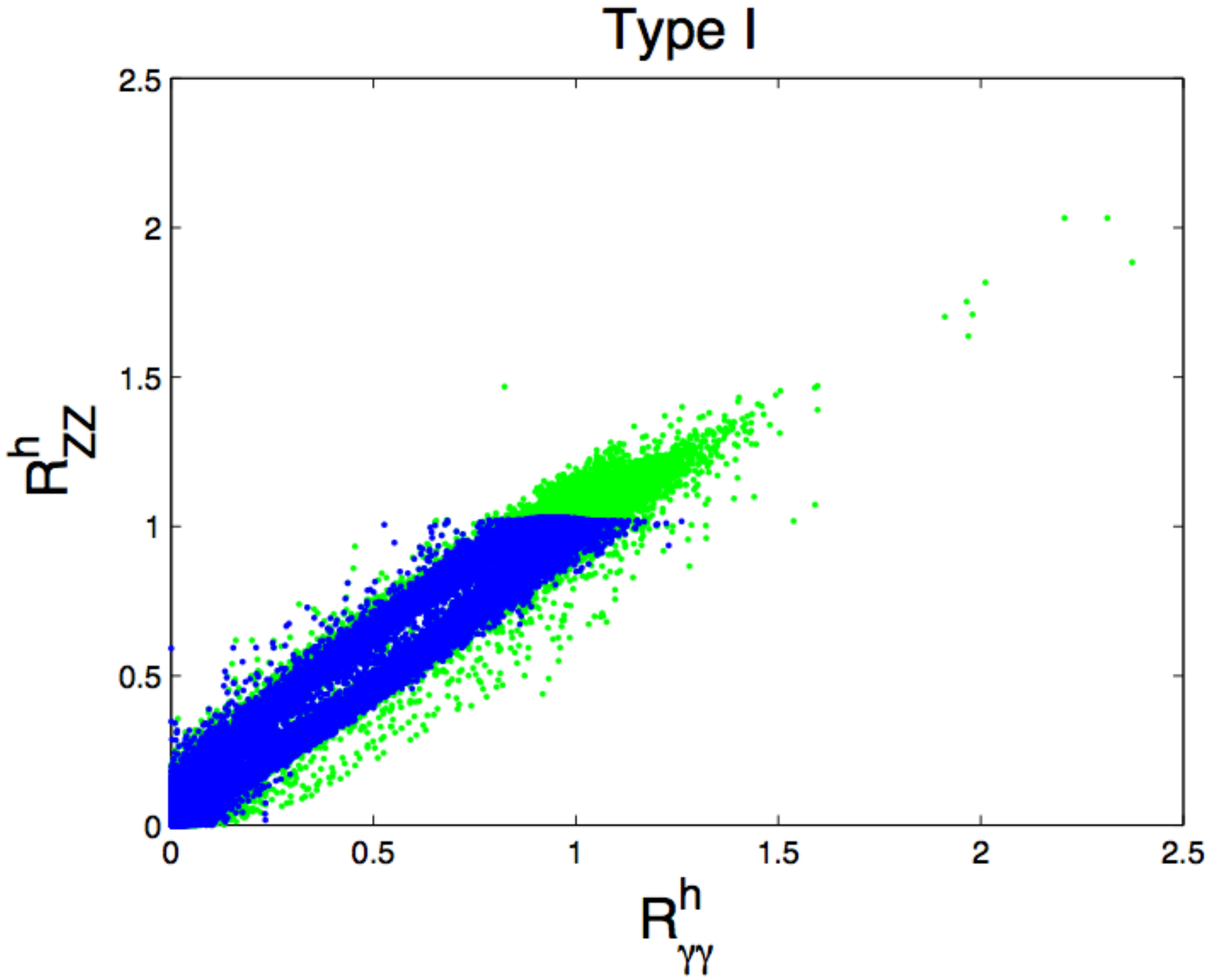}}
{\epsfysize=7cm
\epsfbox{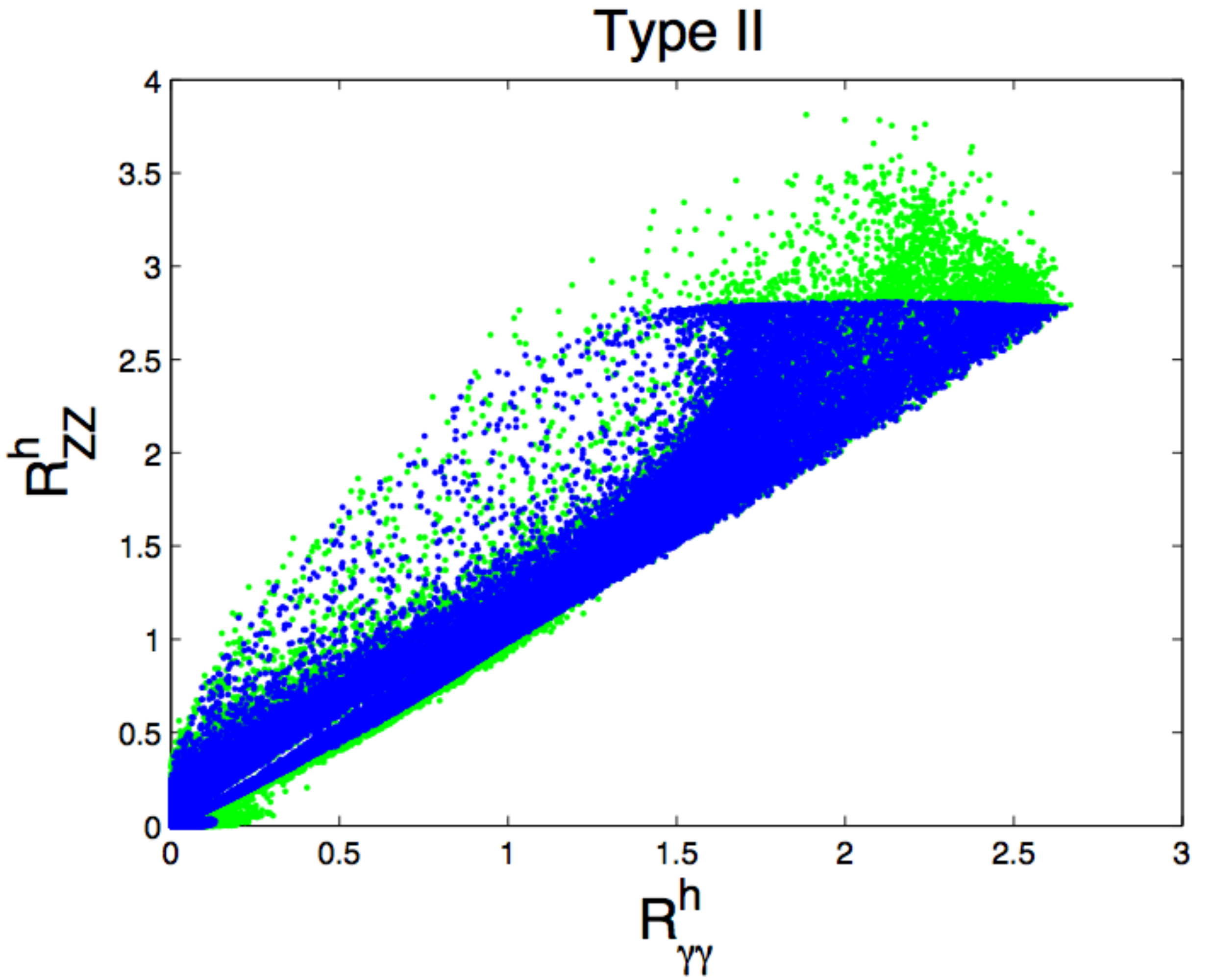}}
\caption{Contributions from chain Higgs production to the rates of $h$ to $ZZ$ and $\gamma\gamma$.
All cuts (B-physics, non observation of heavy scalars, ...) considered. In blue (black) we show
the rates {\em excluding} chain Higgs production. In green (grey) the rates which {\em include}
chain Higgs production.}
\label{fig:chainrates}
\end{figure}
of $h$ to two photons and two $Z$ bosons - meaning, the {\em total} rates, as defined and explained in
eqs.~\eqref{eq:nrat}--~\eqref{eq:nratu}. As one can observe from fig.~\ref{fig:chainrates},
the contributions from chain Higgs production do have an impact on the observables
$R^h_{\gamma\gamma}$ and $R^h_{ZZ}$, by increasing them. This is of course to be expected, since
the chain Higgs process increases the number of $h$ scalars being produced. In particular, we see that
for model type-I chain Higgs production allows one to obtain values of $R^h_{ZZ}$ above unity, something
that would otherwise not be possible for this model. In fact, if one excludes chain Higgs production,
the Higgs production cross section at the LHC is dominated by gluon-gluon fusion which, for model type-I,
is approximately given by
\be
\sigma^{2HDM}(gg\rightarrow h) \simeq \frac{\cos^2\alpha}{\sin^2\beta}\,\sigma^{SM}(gg\rightarrow h).
\ee
On the other hand, the total $h$ width is dominated by the decay $h\rightarrow b\bar{b}$, so that
\be
BR^{2HDM}(h\rightarrow ZZ) \,\simeq\, \frac{\Gamma^{2HDM}(h\rightarrow ZZ)}{\Gamma^{2HDM}(h\rightarrow b\bar{b})}
\,=\,\frac{\sin^2\beta}{\cos^2\alpha}\,\sin^2(\beta - \alpha)\,
\frac{\Gamma^{SM}(h\rightarrow ZZ)}{\Gamma^{SM}(h\rightarrow b\bar{b})}\,,
\ee
and thus
\be
BR^{2HDM}(h\rightarrow ZZ) \,\simeq\,
\frac{\sin^2\beta}{\cos^2\alpha}\,\sin^2(\beta - \alpha)\,BR^{SM}(h\rightarrow ZZ)\, .
\ee
Therefore, under these approximations, the $ZZ$ rate excluding chain Higgs production
is
\be
R^{2HDM}_{ZZ}\,\simeq\, \frac{\sigma^{2HDM}(gg\rightarrow h)\,BR^{2HDM}(h\rightarrow ZZ)}{
\sigma^{SM}(gg\rightarrow h)\,BR^{SM}(h\rightarrow ZZ)} \,=\,\sin^2(\beta - \alpha) \lesssim 1.
\ee
Model type-I therefore predicts values of $R_{ZZ}$ always smaller than about 1. Confirmation of
the higher values measured for this variable by ATLAS could then, in the context of model type-I,
be interpreted as a sign of observation of chain Higgs production. For comparison, it has been
shown~\cite{kane2} that one-loop corrections to the $hZZ$ coupling in the 2HDM are of the order
of less than 5\% for $\cos(\beta - \alpha) \simeq 0$, while for $\cos(\beta - \alpha) \simeq 0.2$
the correction could be as large as 15\%, but {\em negative}. An observed enhancement on $R^h_{ZZ}$
would therefore be difficult to attribute to loop corrections.
\begin{figure}[htb]
\centering
{\epsfysize=7cm
\epsfbox{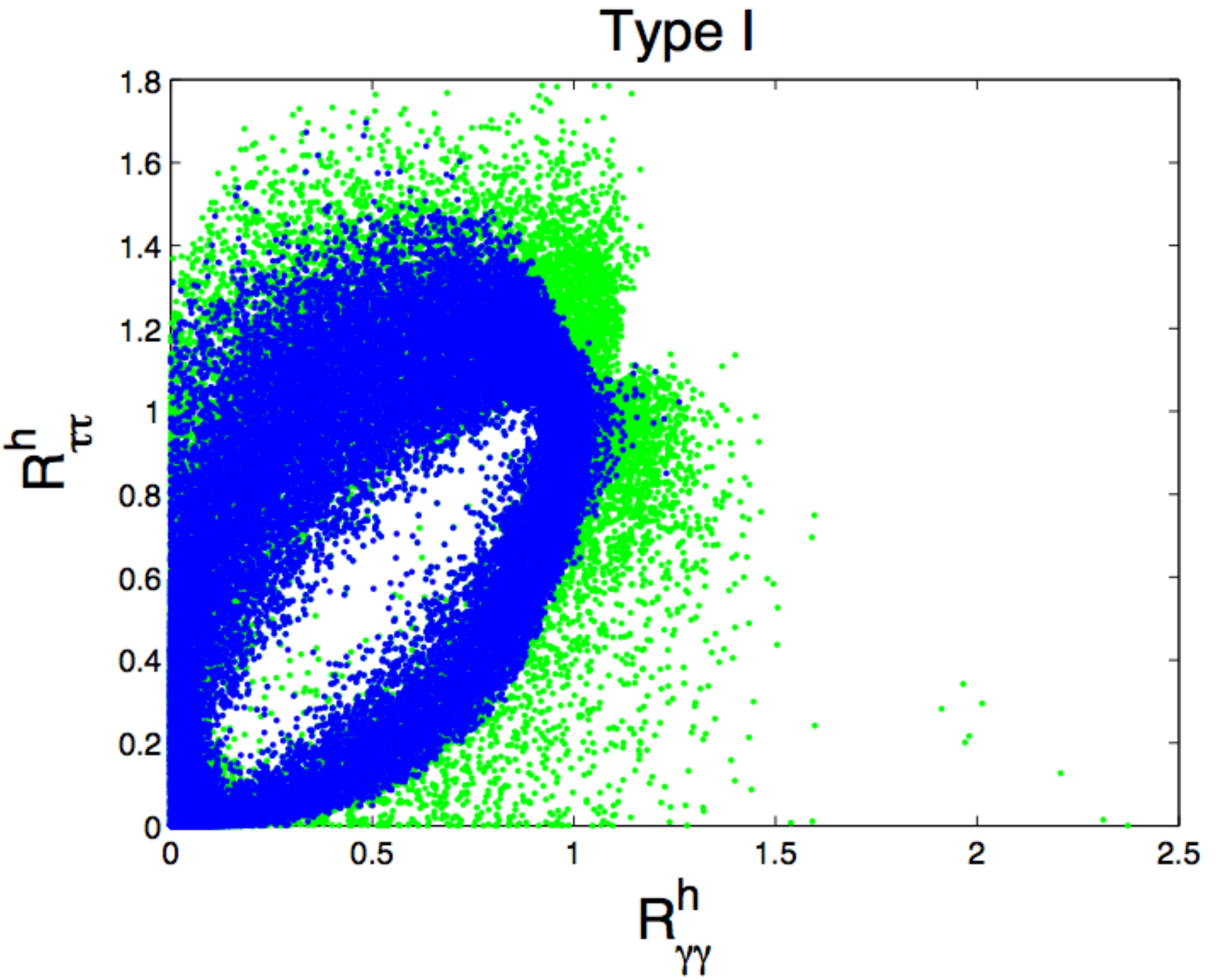}}
{\epsfysize=7cm
\epsfbox{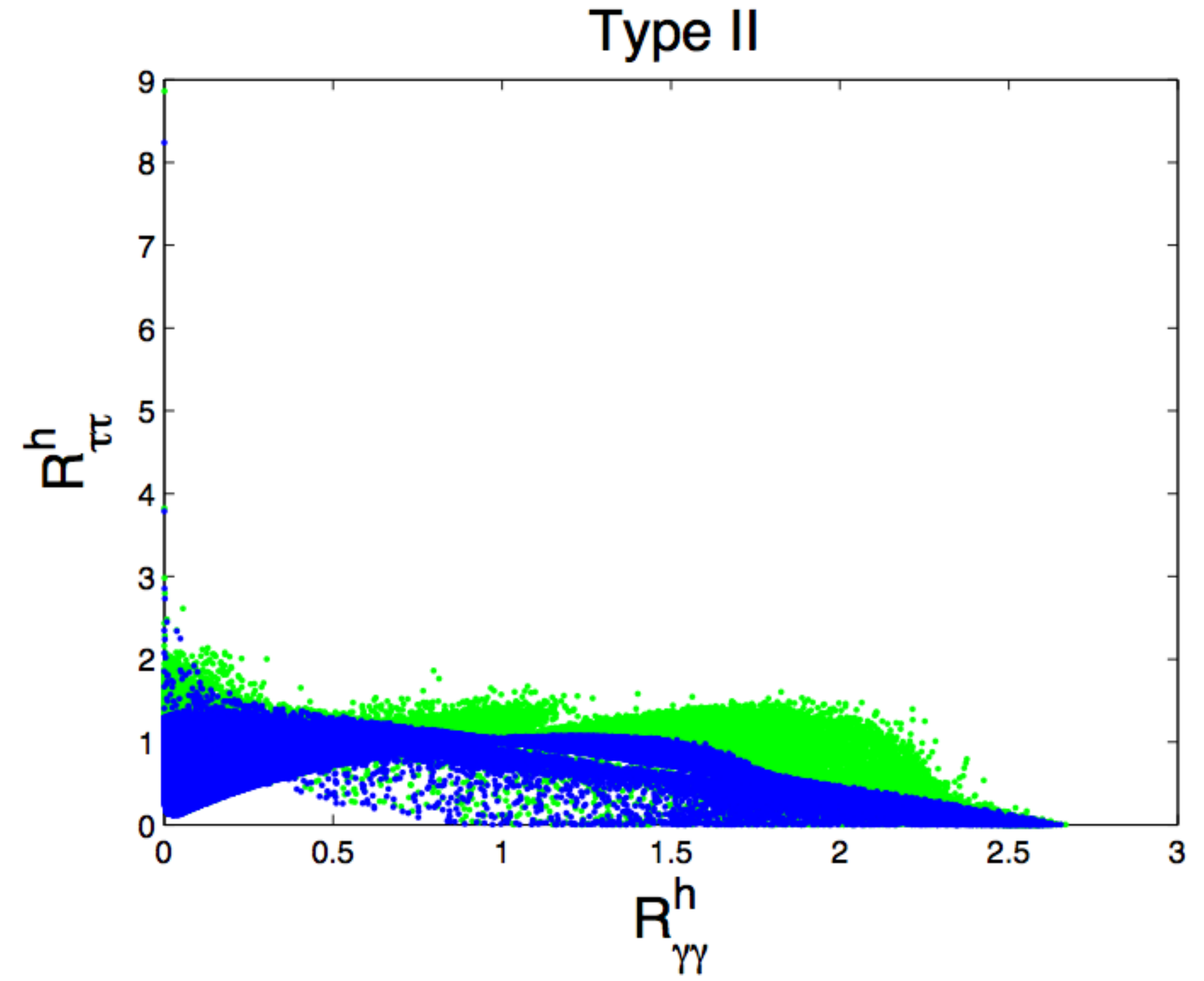}}
\caption{Contributions from chain Higgs production to the rates of $h$ to $\tau\bar{\tau}$ and $\gamma\gamma$.
All constraints (B-physics, non observation of heavy scalars, etc) considered. In blue (black) we show
the rates {\em excluding} chain Higgs production. In green (grey) the rates which {\em include}
chain Higgs production.}
\label{fig:chaintau}
\end{figure}
For completeness, in
fig.~\ref{fig:chaintau} we plot the $h$ rates to $\tau\tau$ and $\gamma\gamma$ with and without
the chain Higgs contributions. Again we see that chain Higgs production tends to increase
the values of these rates, but the effects are not as pronounced as in the previous plots.

However, care must be taken when analysing these extra contributions to the rates $R$. The current
LHC results are the product of a careful and complex dedicated experimental analysis, which
assumes direct
Higgs production. Now, our analysis above did not take into account the kinematic distributions
of any particles in the final states, or intermediate ones. But one obvious place where there will
be a difference in direct or chain Higgs production will be in the transverse momentum ($p_T$)
distributions of the $h$ scalars.~\footnote{We thank S. Dawson, J. Gunion and S. Kraml for discussions
on this subject, after a presentation of an early version of this work in the Scalars 2013 conference.}
{\em Na\"{\i}vely}, we can expect that the $h$ scalars are produced almost at rest in
direct Higgs production, and having a $p_T$ distribution favouring smaller values. However, the chain
Higgs contributions stemming from the decay $H\rightarrow hh$ would generate $h$'s which could have
high $p_T$, which might in turn change considerably the kinematic distributions of their
decay products. Also, chain Higgs production associated with the decays $A\rightarrow Zh$ or
$H^\pm \rightarrow W^\pm h$ would certainly affect the associated production of a scalar $h$ and a
gauge boson.

We are therefore confronted with some very important questions which, to the best of our knowledge,
are thus far unanswered, and require (a) heavy input from our experimentalist colleagues from LHC
and (b) thorough Monte Carlo simulations. In short:
\begin{enumerate}
\item For many choices of parameters, the main contribution to chain Higgs production comes
from the decay $H\rightarrow hh$. Clearly, confirmation of this would come from the identification
of two scalars of masses $\sim$ 125 GeV in the LHC data. However, as explained, it is possible that
these $h$ scalars were not produced at rest and have a $p_T$ distribution quite different from that which
is expected from direct Higgs production. Is it possible to go back to the data already collected and
verify this possibility? One possibility would be to check the presence of two very energetic jets and
two photons, both pairs with invariant masses close to 125 GeV. These requirements should reduce considerably
the QCD backgrounds.
\item Somewhat related to the previous question, does the current experimental analysis performed by the
LHC collaborations even allow for the testing of chain Higgs production? If not, then this is the
perfect time to re-evaluate those choices, since the LHC is undergoing maintenance and will not
begin taking data for another two years.
\item Is it possible that, despite originating in chain Higgs production, the extra $h$ scalars and/or
their decay products might be indistinguishable from those originating in direct Higgs production? This
would seem to certainly be the case for $h$'s being chain-produced via the decays $A\rightarrow Zh$ and
$H^\pm \rightarrow W^\pm h$.
\end{enumerate}

We are confronted with two possibilities, both quite exciting: on the one hand, it is possible that
what is being observed at the LHC already includes signs of a heavy scalar $H$, for instance, but we simply
have not yet performed the correct analyses to detect those new states; on the other hand, it is also possible
that the experimental analyses performed make it at all impossible to test the chain Higgs production
hypothesis.

\section{Conclusions}
\label{sec:conc}

The LHC has provided us with a wealth of data, which confirmed the existence of the Higgs boson.
Within the next years, we should be able to verify whether this new particle conforms perfectly
to its expected SM behaviour or, on the contrary, exhibits properties which indicate the existence
of new physics. One of the simplest extensions of the SM is the 2HDM, which predicts three other
scalars. We have analysed what the current LHC results can tell us about these new scalars. In the mass
scan on the $WW$, $ZZ$, $\tau\bar{\tau}$ and $\gamma\gamma$ channels, non-observation of any excesses
above the expected backgrounds allow us to constrain the 2HDM parameter space. However, there is still
plenty of parameter space available within the model to comply with those restrictions and it is simple, and
in fact expected, for the 2HDM neutral scalars $A$ and $H$ to elude such bounds: $A$, being a pseudoscalar,
does not couple (at tree-level) to $W$'s or $Z$'s, and $H$ is expected to couple very weakly to
gauge bosons, since the CP-even lighter scalar $h$ seems to couple to them with SM-like strength
(as would occur in the decoupling limit). The $\tau\bar{\tau}$ data also excludes a large chunk
of parameter space - mostly for large values of $\tan\beta$, due to the enhanced pseudoscalar
production in that region. The $\gamma\gamma$ signal is also promising if one can obtain exclusion
bounds from LHC for large values of $A$ and $H$ masses.

Another possibility which arises in the 2HDM is that the heavier scalars might be undetectable in the
``usual" channels, because they decay mostly to the lighter scalar $h$. In fact, it is easy to find
vast regions of 2HDM parameter space where the decays $H\rightarrow hh$, or $A\rightarrow Zh$,
or $H^\pm \rightarrow W^\pm h$, are the dominant ones, occurring with branching ratios close to 1.
Under such circumstances, the only way to
detect the heavier scalars would be through $h$ itself. These chain decays of the heavier scalars would
contribute to the lightest Higgs production at the LHC, and we have shown that they might enhance its rates
to $ZZ$, $WW$, $\gamma\gamma$ or $\tau\tau$. Chain Higgs production could therefore be already in effect at
the LHC, and we are seeing signs of the presence of the heavy 2HDM scalars without knowing it.

The unequivocal sign of chain Higgs production would be the identification of two $h$ scalars
being produced at the same time, compatible with being the result of the decay of an $H$ scalar.
This tantalizing possibility can either be indistinguishable from direct Higgs production
with the current experimental analyses, or, even more excitingly, may already be possible to
detect using the current data, requiring that a completely new analysis be performed. In either case,
Monte Carlo simulation, beyond the scope of the current work, is necessary to verify these
possibilities. Also, the experimental search for double Higgs production (within the SM) is not
yet conclusive; when more data is available it may put constraints on some of the channels
we have looked into.

Finally, the issue of chain Higgs production is not exclusive to the 2HDM. That possibility could also be
present in SUSY models, and in fact in any multi-Higgs doublet models. Any thorough analysis of what the
current LHC data can tell us about chain Higgs production is therefore also relevant for studies of
those other models.

\begin{acknowledgments}
We thank R. Gon\c{c}alo, F. Maltoni and  P. Conde Mui{\~n}o for discussions.
We are  grateful for the hospitality of Toyama University, Japan,
where part of this work was carried out.
The works of P.M.F. and R.S. are supported in part by the Portuguese
\textit{Funda\c{c}\~{a}o para a Ci\^{e}ncia e Tecnologia} (FCT)
under contract PTDC/FIS/117951/2010 and by PEst-OE/FIS/UI0618/2011
and by a FP7 Reintegration Grant, number PERG08-GA-2010-277025.
The works of A.A.,  P.M.F. and R.S. are partially supported by a Bilateral Agreement FCT-CNRST Grant.
\end{acknowledgments}


\begin{thebibliography}{99}
%
\bibitem{:2012gk}
G.~Aad {\it et al.}  [ATLAS Collaboration],
Phys.\ Lett.\ B \textbf{716}, 1 (2012)
[arXiv:1207.7214 [hep-ex]].

\bibitem{:2012gu}
S.~Chatrchyan \textit{et al.}  [CMS Collaboration],
Phys.\ Lett.\ B \textbf{716}, 30 (2012)
[arXiv:1207.7235 [hep-ex]].

\bibitem{Lee:1973iz}
 T.~D.~Lee,
 Phys.\ Rev.\  D {\bf 8} (1973) 1226.

\bibitem{Branco:2011iw}
  G.~C.~Branco, P.~M.~Ferreira, L.~Lavoura, M.~N.~Rebelo, M.~Sher and J.~P.~Silva,
  Phys.\ Rept.\  {\bf 516}, 1 (2012)
  [arXiv:1106.0034 [hep-ph]].

\bibitem{Ivanov:2006yq}
 I.~P.~Ivanov,
 Phys.\ Rev.\  D {\bf 75} (2007) 035001
 [Erratum-ibid.\  D {\bf 76} (2007) 039902]
 [arXiv:hep-ph/0609018].

\bibitem{Ivanov:2007de}
 I.~P.~Ivanov,
 Phys.\ Rev.\  D {\bf 77} (2008) 015017
 [arXiv:0710.3490 [hep-ph]].

\bibitem{Barroso:2007rr}
  A.~Barroso, P.~M.~Ferreira and R.~Santos,
  Phys.\ Lett.\ B {\bf 652} (2007) 181
  [hep-ph/0702098 [HEP-PH]].

 \bibitem{Ferreira:2004yd}
 P.~M.~Ferreira, R.~Santos and A.~Barroso,
 Phys.\ Lett.\  B {\bf 603} (2004) 219
 [Erratum-ibid.\  B {\bf 629} (2005) 114]
 [arXiv:hep-ph/0406231].

\bibitem{Barroso:2005sm}
 A.~Barroso, P.~M.~Ferreira and R.~Santos,
 Phys.\ Lett.\  B {\bf 632} (2006) 684
 [arXiv:hep-ph/0507224].

 \bibitem{Barroso:2012mj}
  A.~Barroso, P.~M.~Ferreira, I.~P.~Ivanov, R.~Santos and J.~P.~Silva,
  Eur.\ Phys.\ J.\ C {\bf 73}, 2537 (2013)
  [arXiv:1211.6119 [hep-ph]].

 \bibitem{Barroso:2013awa}
  A.~Barroso, P.~M.~Ferreira, I.~P.~Ivanov and R.~Santos,
  JHEP {\bf 1306} (2013) 045
  [arXiv:1303.5098 [hep-ph]].

\bibitem{Glashow:1976nt}
 S.~L.~Glashow and S.~Weinberg,
 Phys.\ Rev.\  D {\bf 15} (1977) 1958.

\bibitem{Paschos:1976ay}
 E.~A.~Paschos,
 Phys.\ Rev.\  D {\bf 15} (1977) 1966.

 \bibitem{Branco:1996bq}
 G.~C.~Branco, W.~Grimus and L.~Lavoura,
 Phys.\ Lett.\  B {\bf 380} (1996) 119
 [arXiv:hep-ph/9601383].

\bibitem{inert}
 E.~Ma,
 Phys.\ Rev.\  D {\bf 73} (2006) 077301
 [hep-ph/0601225];

 R.~Barbieri, L.~J.~Hall, and V.~S.~Rychkov,
 Phys.\ Rev.\  D {\bf 74} (2006) 015007
 [hep-ph/0603188];

 R.~Barbieri and A.~Strumia,
 hep-ph/0007265;

  H.~Mart\'\i nez, A.~Melfo, F.~Nesti, and G.~Senjanovi\'c,
  Phys.\ Rev.\ Lett.\ {\bf 106} (2011) 191802
  [arXiv:1101.3796 [hep-ph]];

 E.~Ma,
 Mod.\ Phys.\ Lett.\  A {\bf 21} (2006) 1777
 [hep-ph/0605180];

 D.~Majumdar and A.~Ghosal,
 Mod.\ Phys.\ Lett.\  A {\bf 23} (2008) 2011
 [hep-ph/0607067];

  L.~L.~Honorez, E.~Nezri, J.~F.~Oliver, and M.~H.~G.~Tytgat,
  JCAP {\bf 0702} (2007) 028
  [hep-ph/0612275];

 N.~Sahu and U.~Sarkar,
 Phys.\ Rev.\  D {\bf 76} (2007) 045014
 [hep-ph/0701062];

 M.~Gustafsson, E.~Lundstrom, L.~Bergstrom, and J.~Edsj\"o,
 Phys.\ Rev.\ Lett.\  {\bf 99} (2007) 041301
 [arXiv:astro-ph/0703512];

 M.~Lisanti and J.~G.~Wacker,
 arXiv:0704.2816 [hep-ph];

 T.~Hambye, F.~S.~Ling, L.~L.~Honorez, and J.~Rocher,
 JHEP {\bf 0907} (2009) 090
 [Erratum {\it ibid.}\  {\bf 1005} (2010) 066]
 [arXiv:0903.4010 [hep-ph]];

  A.~Melfo, M.~Nemevsek, F.~Nesti, G.~Senjanovi\'c, and Y.~Zhang,
  arXiv:1105.4611 [hep-ph];

  I.~F.~Ginzburg, K.~A.~Kanishev, M.~Krawczyk, and D.~Sokolowska,
  Phys.\ Rev.\ D {\bf 82} (2010) 123533
  [arXiv:1009.4593 [hep-ph]];

A.~Arhrib, Y.~-L.~S.~Tsai, Q.~Yuan and T.~-C.~Yuan,
  arXiv:1310.0358 [hep-ph].

\bibitem{heavyh}
  P.~M.~Ferreira, R.~Santos, M.~Sher and J.~P.~Silva,
  Phys.\ Rev.\  D {\bf 85}, 035020 (2012)
  [arXiv:1201.0019 [hep-ph]];

  G.~Belanger, U.~Ellwanger, J.~F.~Gunion, Y.~Jiang, S.~Kraml and J.~H.~Schwarz,
  JHEP {\bf 1301} (2013) 069
  [arXiv:1210.1976 [hep-ph]].

\bibitem{mix}
A.~Barroso, P.~M.~Ferreira, R.~Santos and J.~P.~Silva,
  Phys.\ Rev.\ D {\bf 86} (2012) 015022
  [arXiv:1205.4247 [hep-ph]].

\bibitem{Chen:2013rba}
  C.~-Y.~Chen, S.~Dawson and M.~Sher,
  Phys.\ Rev.\ D {\bf 88} (2013) 015018
  [arXiv:1305.1624 [hep-ph]].

\bibitem{chdm}
  I.~F.~Ginzburg, M.~Krawczyk and P.~Osland,
[hep-ph/0211371];

  W.~Khater and P.~Osland,
Nucl.\ Phys.\ B \textbf{661}, 209 (2003)  [hep-ph/0302004];


  A.~Arhrib, E.~Christova, H.~Eberl and E.~Ginina,
JHEP {\bf 1104}, 089 (2011)
[arXiv:1011.6560 [hep-ph]].

\bibitem{othermodels}
V.~D.~Barger, J.~L.~Hewett and R.~J.~N.~Phillips,
 Phys. \ Rev.\  D {\bf 41} (1990) 3421;

 M.~Aoki, S.~Kanemura, K.~Tsumura and K.~Yagyu,
 Phys.\ Rev.\  D {\bf 80} (2009) 015017.

\bibitem{vac1}
  N.G.~Deshpande and E.~Ma,
  Phys.\ Rev.\  D {\bf 18} (1978) 2574.

\bibitem{unitarity}
S.~Kanemura, T.~Kubota and E.~Takasugi,
Phys.\ Lett.\  B {\bf 313} (1993)  155;

A.G.~Akeroyd, A.~Arhrib and E.M.~Naimi,
  Phys.\ Lett.\  B {\bf 490} (2000)  119.

\bibitem{Peskin:1991sw}
  M.E.~Peskin and T.~Takeuchi,
  Phys.\ Rev.\ D {\bf 46}, 381 (1992).

\bibitem{STHiggs}
 H.E.~Haber,
  ``Introductory Low-Energy Supersymmetry,'' in
\textit{Recent directions in particle theory: from superstrings and black holes to the standard model}, Proceedings of
the Theoretical Advanced Study Institute (TASI 92), Boulder, CO, 1--26
June 1992, edited by J.~Harvey and J.~Polchinski (World Scientific
Publishing, Singapore, 1993) pp.~589--688;

  C.~D.~Froggatt, R.~G.~Moorhouse and I.~G.~Knowles,
  Phys.\ Rev.\  D {\bf 45}, 2471 (1992);

 W.~Grimus, L.~Lavoura, O.~M.~Ogreid and P.~Osland,
  Nucl.\ Phys.\ B {\bf 801}, 81 (2008)
  [arXiv:0802.4353 [hep-ph]];

  H.E.~Haber and D.~O'Neil,
  Phys.\ Rev.\ D {\bf 83}, 055017 (2011)
  [arXiv:1011.6188 [hep-ph]].

\bibitem{LEP2013}
  G.~Abbiendi {\it et al.}  [ALEPH and DELPHI and L3 and OPAL and The LEP working group for Higgs boson searches Collaborations],
  [arXiv:1301.6065 [hep-ex]].

\bibitem{BB}
  A.~Denner, R.J.~Guth, W.~Hollik and J.H.~Kuhn,
  Z.\ Phys.\ C {\bf 51}, 695 (1991).

  H.E.~Haber and H.E.~Logan,
  Phys.\ Rev.\ D {\bf 62}, 015011 (2000)
  [hep-ph/9909335].

The ALEPH, CDF,  D0, DELPHI, L3, OPAL, SLD Collaborations, the LEP Electroweak Working Group, the Tevatron Electroweak Working Group, and the SLD electroweak and heavy flavour Groups,
  arXiv:1012.2367 [hep-ex].

  F.~Mahmoudi and O.~Stal,
  Phys.\ Rev.\ D {\bf 81}, 035016 (2010)
  [arXiv:0907.1791 [hep-ph]];

  M.~Baak, M.~Goebel, J.~Haller, A.~Hoecker, D.~Ludwig, K.~Moenig, M.~Schott and J.~Stelzer,
  Eur.\ Phys.\ J.\ C {\bf 72}, 2003 (2012)
  [arXiv:1107.0975 [hep-ph]].

M.~Baak, M.~Goebel, J.~Haller, A.~Hoecker, D.~Kennedy, R.~Kogler, K.~Moenig, M.~Schott and J.~Stelzer,
  arXiv:1209.2716 [hep-ph];

  A.~Wahab El Kaffas, P.~Osland, O.~M.~Ogreid,
  Phys.\ Rev.\ D {\bf 76}, 095001 (2007)
  [arXiv:0706.2997 [hep-ph]];

  L.~Basso, A.~Lipniacka, F.~Mahmoudi, S.~Moretti, P.~Osland, G.~M.~Pruna, M.~Purmohammadi,
  JHEP {\bf 1211}, 011 (2012)
  [arXiv:1205.6569 [hep-ph]].

\bibitem{BB2}
  D.~Asner {\it et al.}  [Heavy Flavor Averaging Group Collaboration],
  arXiv:1010.1589 [hep-ex];

T.~Hermann, M.~Misiak and M.~Steinhauser,
  JHEP {\bf 1211} (2012) 036.

\bibitem{ATLASICHEP}
  The ATLAS collaboration, ATLAS-CONF-2013-090.
  G.~Aad {\it et al.}  [ATLAS Collaboration],
  JHEP {\bf 1206} (2012) 039
  [arXiv:1204.2760 [hep-ex]].

\bibitem{CMSICHEP}
  S.~Chatrchyan {\it et al.}  [CMS Collaboration],
  JHEP {\bf 1207} (2012) 143
  [arXiv:1205.5736 [hep-ex]];

\bibitem{mssmhiggs}
 S.~Schael {\it et al.}  [ALEPH and DELPHI and L3 and OPAL and LEP Working Group for Higgs Boson Searches Collaborations],
  Eur.\ Phys.\ J.\ C {\bf 47}, 547 (2006)
  [hep-ex/0602042].

\bibitem{Lees:2012xj}
  J.P.~Lees {\it et al.}  [BaBar Collaboration],
  Phys.\ Rev.\ Lett.\  {\bf 109}, 101802 (2012)
  [arXiv:1205.5442 [hep-ex]].

\bibitem{LHCHiggs}
https://twiki.cern.ch/twiki/bin/view/LHCPhysics/CrossSectionsFigures\#Higgs\_production\_cross\_sections


\bibitem{Harlander:2003ai}
R.V.~Harlander and W.B.~Kilgore,
Phys.\ Rev.\ D {\bf 68}, 013001 (2003)
[hep-ph/0304035].

\bibitem{Spira:1995mt}
M.~Spira,
arXiv:hep-ph/9510347.

\bibitem{Djouadi:1999rca}
A.~Djouadi, W.~Kilian, M.~Muhlleitner and P.~M.~Zerwas,
Eur.\ Phys.\ J.\ C {\bf 10}, 45 (1999)
[hep-ph/9904287].

\bibitem{Arhrib:2009hc}
A.~Arhrib, R.~Benbrik, C.~-H.~Chen, R.~Guedes and R.~Santos,
JHEP {\bf 0908}, 035 (2009)
[arXiv:0906.0387 [hep-ph]].

\bibitem{Huitu:2010ad}
K.~Huitu, S.~Kumar Rai, K.~Rao, S.~D.~Rindani and P.~Sharma,
JHEP {\bf 1104}, 026 (2011)
[arXiv:1012.0527 [hep-ph]].

\bibitem{Aoki:2011wd}
M.~Aoki, R.~Guedes, S.~Kanemura, S.~Moretti, R.~Santos and K.~Yagyu,
Phys.\ Rev.\ D {\bf 84}, 055028 (2011)
[arXiv:1104.3178 [hep-ph]].

\bibitem{Ferreira:2013qua}
P.~M.~Ferreira, R.~Santos, M.~Sher and J. P.~Silva,
arXiv:1305.4587 [hep-ph];
A.~Barroso, P.~M.~Ferreira, R.~Santos, M.~Sher and J. P.~Silva,
arXiv:1304.5225 [hep-ph].

\bibitem{atlas_zzww}
G.~Aad {\it et al.}  [ATLAS Collaboration], ATLAS-CONF-2013-067;

G.~Aad {\it et al.}  [ATLAS Collaboration],
  Phys.\ Lett.\ B {\bf 710} (2012) 383
  [arXiv:1202.1415 [hep-ex]].

\bibitem{cms_zzww}
S.~Chatrchyan {\it et al.} [CMS Collaboration],
  Eur.\ Phys.\ J.\ C {\bf 73} (2013) 2469
  [arXiv:1304.0213 [hep-ex]].

\bibitem{zzback}
  J.~M.~Campbell, R.~K.~Ellis and C.~Williams,
  JHEP {\bf 1110} (2011) 005
  [arXiv:1107.5569 [hep-ph]];

   N.~Kauer and G.~Passarino,
  JHEP {\bf 1208} (2012) 116
  [arXiv:1206.4803 [hep-ph]];

  D.~Buarque Franzosi, F.~Maltoni and C.~Zhang,
  Phys.\ Rev.\ D {\bf 87} (2013) 053015
  [arXiv:1211.4835 [hep-ph]].
\bibitem{atlas_tautau}
  G.~Aad {\it et al.}  [ATLAS Collaboration],
  JHEP {\bf 1302} (2013) 095
  [arXiv:1211.6956 [hep-ex]].

\bibitem{tau}
A.~Arhrib, C.~-W.~Chiang, D.~K.~Ghosh and R.~Santos,
 Phys.\ Rev.\ D {\bf 85} (2012) 115003
 [arXiv:1112.5527 [hep-ph]].

\bibitem{atlas_phph}
G.~Aad {\it et al.}  [ATLAS Collaboration],
  Phys.\ Rev.\ Lett.\  {\bf 108} (2012) 111803
  [arXiv:1202.1414 [hep-ex]].

\bibitem{cms_phph}
S.~Chatrchyan {\it et al.}  [CMS Collaboration],CMS-HIG-13-016.

\bibitem{DKZ}
  A.~Djouadi, J.~Kalinowski and P.~M.~Zerwas,
  Z.\ Phys.\ C {\bf 70} (1996) 435
  [hep-ph/9511342].

\bibitem{kane2}
S.~Kanemura, Y.~Okada, E.~Senaha and C.~-P.~Yuan,
  Phys.\ Rev.\ D {\bf 70} (2004) 115002
  [hep-ph/0408364].


\end{thebibliography}
\end{document}